\documentclass[iop]{emulateapj}
\usepackage{apjfonts}
\usepackage{graphicx}
\usepackage{color}

\usepackage{epstopdf}
\usepackage{amsmath}

\def\ltsima{$\; \buildrel < \over \sim \;$}
\def\simlt{\lower.5ex\hbox{\ltsima}}
\def\gtsima{$\; \buildrel > \over \sim \;$}
\def\simgt{\lower.5ex\hbox{\gtsima}}

\newcommand\lsim{\mathrel{\rlap{\lower4pt\hbox{\hskip1pt$\sim$}}
\raise1pt\hbox{$<$}}}
\newcommand\gsim{\mathrel{\rlap{\lower4pt\hbox{\hskip1pt$\sim$}}
\raise1pt\hbox{$>$}}}

\usepackage{color}

\shorttitle{EKL for Outer Orbit Test Particle }
\shortauthors{S. Naoz et al}

\begin{document}

\title{The Eccentric Kozai-Lidov mechanism for Outer Test Particle}

\author{  Smadar Naoz$^1$  \& Gongjie Li$^2$ \\ Macarena  Zanardi$^{3,4}$, Gonzalo Carlos  de El\'{\i}a$^{3,4}$, Romina P.~Di Sisto$^{3,4}$ }
\altaffiltext{1}{Department of Physics and Astronomy, University of California, Los Angeles, CA 90095, USA}
\altaffiltext{2}{Harvard Smithsonian Center for Astrophysics, Institute for Theory and Computation, 60 Garden Street,
Cambridge, MA 02138, USA}
\altaffiltext{3}{ Instituto de Astrof\'{\i}sica de La Plata, CCT La Plata-CONICET-UNLP Paseo del Bosque S/N (1900), La Plata, Argentina}
\altaffiltext{4}{Facultad de Ciencias Astron\'{o}micas y Geof\'{\i}sicas, Universidad Nacional de La Plata Paseo del Bosque S/N (1900), La Plata, Argentina}
\email{snaoz@astro.ucla.edu}

\begin{abstract}

The secular approximation of the hierarchical three body  systems has been proven to be very useful in addressing many astrophysical systems, from planets, stars to black holes. In such a system two objects are on a tight orbit, and the tertiary is on a much wider orbit.   Here we study the dynamics of a system by taking the tertiary mass to zero and solve the hierarchical three body system up to the octupole level of approximation. We find a rich dynamics that the outer orbit undergoes due to gravitational perturbations from the inner binary.  The nominal result of the precession of the nodes is mostly limited for the lowest order of approximation, however, when the octupole-level of approximation is introduced the system becomes chaotic, as expected, and the tertiary oscillates below and above $90^\circ$, similarly to the non-test particle flip behavior \citep[e.g.,][]{Naoz16}. We provide the Hamiltonian of the system and investigate the dynamics of the system from the quadrupole to the octupole level of approximations. We also analyze  the chaotic and quasi-periodic orbital evolution by studying the surfaces of sections.  Furthermore, including general relativity, we show case the long term evolution of individual debris disk particles under the influence of a far away interior eccentric planet. We show that this dynamics can naturally result in  retrograde objects and a puffy disk after a long timescale evolution (few Gyr) for initially aligned configuration.
\end{abstract}

%\keywords{stars: binaries: close, stars: black holes, evolution, kinematics and dynamics, X-rays: binaries} 

\maketitle

\section{INTRODUCTION}

The hierarchical three body secular dynamics has been studied extensively in the literature and was shown to be very effective  in addressing different astrophysical phenomena  \citep[see for review][and reference therein]{Naoz16}. In this hierarchical  setting the {\it inner binary}  is orbited by a third body on a much wider orbit,  the {\it outer binary}, such that the secular approximation can be applied (i.e., phase averaged, long-term interaction). % In other words, the interactions between two  non-resonant orbits is equivalent to treating the two orbits as massive wires, where the line-density is inversely proportional to orbital velocity. 
  The gravitational potential is then expanded in semi-major axis ratio  ($a_1/a_2$, which, in this approximation, remains constant), where  $a_1$ ($a_2$) is the semi-major axis of the inner (outer) body \citep{Kozai, Lidov}.  This ratio  is a small parameter due to the hierarchical configuration. The lowest order of  approximation, which is proportional to $(a_1/a_2)^2$  is called the quadrupole-level. 

Most of these studies focus {\bf on} the gravitational perturbations that a far away perturber exerts on the inner binary. 
In early studies of high-inclination secular perturbations
\citep{Kozai,Lidov}, the outer orbit was assumed to be circular and it was assumed  that  one of the inner binary members is a massless test particle. 
 In this situation, the
component of the inner orbit's angular momentum along the z-axis (which is set to be parallel to the total angular momentum, i.e., the invariable plane) is
conserved, and the lowest order of the approximation, the quadrupole approximation, is valid. However, relaxing either one of these assumptions leads to qualitative different behavior \citep[e.g.,][]{Naoz11,LN,Katz+11}. Considering systems beyond the test particle approximation, or a circular orbit, requires the next level of approximation, called the octupole--level of approximation, which is proportional  to $(a_1/a_2)^3$ \citep[e.g.][]{Har68,Har69,Ford00,Bla+02}.  In the octupole level of approximation, the inner orbit eccentricity may reach extreme values \citep{Ford00,Naoz+11sec,Li+13,Tey+13}.  In addition, the inner orbit  can flip its orientation, with respect to the total angular momentum (i.e., z-axis), from prograde to retrograde \citep{Naoz11}.  

Here we study the secular evolution of a far away test particle orbiting an inner massive binary. In this case, the inner orbit is fixed, and effectively carries all of the annular momentum of the system, while the outer orbit undergoes a dynamical evolution. This situation has large range of applications from the gravitational perturbations of binary super massive black holes on the surrounding stellar distribution to the effects of planetary orbits on debris disks,  Oort cloud and  trans-neptunian objects.  
From N-body simulations, \citet{Zanardi+17} analyzed the long term evolution of test particles in the presence of an interior eccentric planet. Such an study produces particles on prograde and retrograde orbits, as well as particles whose orbital plane flips from prograde to retrograde and back again along their evolution.

% Zanardi et al (2017) numerical results of the long term evolution of test particles in the presence of  interior eccentric planet(s) found retrograde test particles. Motivated by these results we find the general dynamics of an outer test particle in a hierarchical triple system.

  We note that \citet{Ziglin75}  investigated the oscillations of an outer circumbinary planet in the context of the restricted elliptical three body problem.   Later \citet{Verrier+09} and  \citet{Far+10}  studied  the stability of high inclined planet around in this situation using a combination of numerical and perturbation theory up to the quadruple level of approximation approaches \citep[see also,][]{Li+14cir,Marcos+15}. 
 {Furthermore}, \citet{Gallardo06} and \citet{Gallardo+12} studied the effects of the Kozai-Lidov for  trans-neptunian objects near mean motion resonance with Neptune. However, here, we do not allow for mean motion resonances to allow for the double averaging process \citep[see][appendix A2 for the canonical transformation which describes the averaging process]{Naoz+11sec}.We provide a general treatment for the outer-test particle case, up to the octupole--level of approximation in the secular theory. 

The paper is organized as follow: We begin by describing the outer test particle Hamiltonian and  equations of motion (\S \ref{sec:EOM}), and continue to discuss the quadrupole-level of approximation (\S \ref{sec:quad}) where we also drive the relevant timescales, and then we study the role  of the octupole-level of approximations, and provide surface of section maps (\S  \ref{sec:oct}).  We also discuss the role of general relativity precession in  \S \ref{sec:GR}. We then consider one study case in the form of the long term evolution debris disk particles in \S \ref{sec:Kuip}. Finally, we offer our discussions in \S \ref{sec:dis}.
 
\section{The Equations of Motion}\label{sec:EOM}
We solve the orbit of an {\it exterior} massless test particle to an eccentric planet ($m_2$), both orbiting a star ($m_1$), including only secular interactions expanded to octupole order.  The  planet is on a fixed eccentric orbit (i.e., $e_1={\rm const}$) and the outer particle's orbit is specified by four variables:
\begin{equation}
{e_2,\omega_2,\theta,\Omega_2} \ ,
\end{equation}
where $e_2$ is the test particle eccentricity,  $\theta=\cos i$ and $i$ is the inclination of the test particle with respect to the inner orbit, and $\omega_2$ and $\Omega_2$ are the argument of periapse and longitude of ascending node of the outer orbit, relative to the inner planet's periapse \citep{MD00}. Specifically, We set $\varpi_1 = 0$, see Appendix \ref{App:Ham} for the coordinate transformation. We kept the subscript ``2" in $\omega_2$ and $\Omega_2$  for consistency with the comparable masses treatments. %, but it does not implies that the 
%Note that $\omega$ and $\Omega$ are denoted usually as $\omega_2$ and $\Omega_2$ in comparable masses treatments.
From $e_2$ and $\theta$ we can define the canonical specific momenta 
\begin{eqnarray}
J_2&=&\sqrt{1-e_2^2} \\
J_{2,z}&=&\theta \sqrt{1-e_2^2}
\end{eqnarray}

The Hamiltonian for which $m_3\to0$ is 
\begin{equation}
f=f_{\rm quad} + \epsilon_M f_{\rm oct} \ ,
\end{equation}
where 
\begin{equation}
 \epsilon_M =\frac{m_1-m_2}{m_1+m_2} \frac{a_1}{a_2}\frac{e_2}{1-e_2^2}
\end{equation}
and
\begin{equation}\label{eq:fquad}
f_{\rm quad}  = \frac{ (2+3e_1^2)(3\theta^2-1)+15e_1^2(1-\theta^2)\cos(2\Omega_2)}{(1-e_2^2)^{3/2}} \ ,
\end{equation}
\begin{eqnarray}\label{eq:foct}
f_{\rm oct}  &= & \frac{ 15 e_1}{4 (1-e_2^2)^{3/2}} \bigg[ 10(1-e_1^2)\theta (1-\theta^2) \sin \omega_2\sin\Omega_2 \nonumber  \\ 
&+& \frac{1}{2}\{ 2+19e_1^2-5(2+5e_1^2)\theta^2-35e_1^2(1-\theta^2)\cos(2\Omega_2) \} \nonumber \\
&\times& (\theta \sin\omega_2 \sin\Omega_2 -\cos\omega_2\cos\Omega_2) \bigg]
\end{eqnarray}
Note that unlike the {\it inner} test particle approximation $ \epsilon_M$ is not constant during the motion, and the constant parameter during the evolution is:
\begin{equation}
\delta=\frac{m_1-m_2}{m_1+m_2} \frac{a_1}{a_2}e_1
\end{equation}
and the Hamiltonian up to the octupole level of approximation   can be defined as:
\begin{equation}
\tilde{f}_{\rm oct}= \frac{e_2}{1-e_2^2}\frac{f_{\rm oct} }{e_1}
\end{equation}
and \begin{equation} f=f_{\rm quad} +  \delta \tilde{f}_{\rm oct} \ . \end{equation}
%In Appendix A, we use the secular octupole Hamiltonian that has been published in the literature to derive the particle's equations of motion. %As we show in Appendix A, although that published Hamiltonian has had its nodes eliminated, one can still use it to derive the full test particle equations of motions. %  the equation that requires the nodes.
We note that $f_{\rm quad}$ has the same functional form as the inner test particle $F_{\rm quad}$ presented in \citet{LN} up to the $(1-e_2^2)^{3/2}$ which is not constant in our case.

The equations of motion may be expressed as partial derivatives of an energy function $f(e_2,\omega_2,\theta,\Omega_2)$
\begin{eqnarray}\label{eq:EOM1}
\frac{dJ_2}{d\tau} &=&  \frac{\partial f}{\partial \omega_2}  \\ 
\frac{dJ_{2,z}}{d\tau} &=&\frac{\partial f}{\partial \Omega_2} \\
\frac{d\omega_2 }{d\tau} &=& \frac{\partial f}{\partial e_2}\frac{J_2}{e_2} +  \frac{\partial f}{\partial \theta}\frac{\theta }{J_2} \\\label{eq:EOM4}
\frac{d\Omega_2 }{d\tau} &=& -  \frac{\partial f}{\partial \theta}\frac{1 }{J_2}
\end{eqnarray} 
where $\tau$ is proportional to the true time (see Eq.~[\ref{eq:time}]).
Unlike inner orbit test particle  $\epsilon_M$ is not constant while $e_1$ is constant (in other words, the angular momentum of the inner orbit is conserved). The equations of motion were tested successfully compared to the general equations of motions, presented in   \citet{Naoz+11sec}. We also test the evolution compared to N-body in Appendix \ref{sec:Nbody}.

 % {\bf prefer to delete "the only thing that is constant here is" since $\delta$ is also a constant} {\bf while $e_1$ is constant }(in other words, the angular momentum of the inner orbit).

\section{quadrupole-level of approximation }\label{sec:quad}
\subsection{General Analysis}
\begin{figure}
\begin{center}
\includegraphics[width=\linewidth]{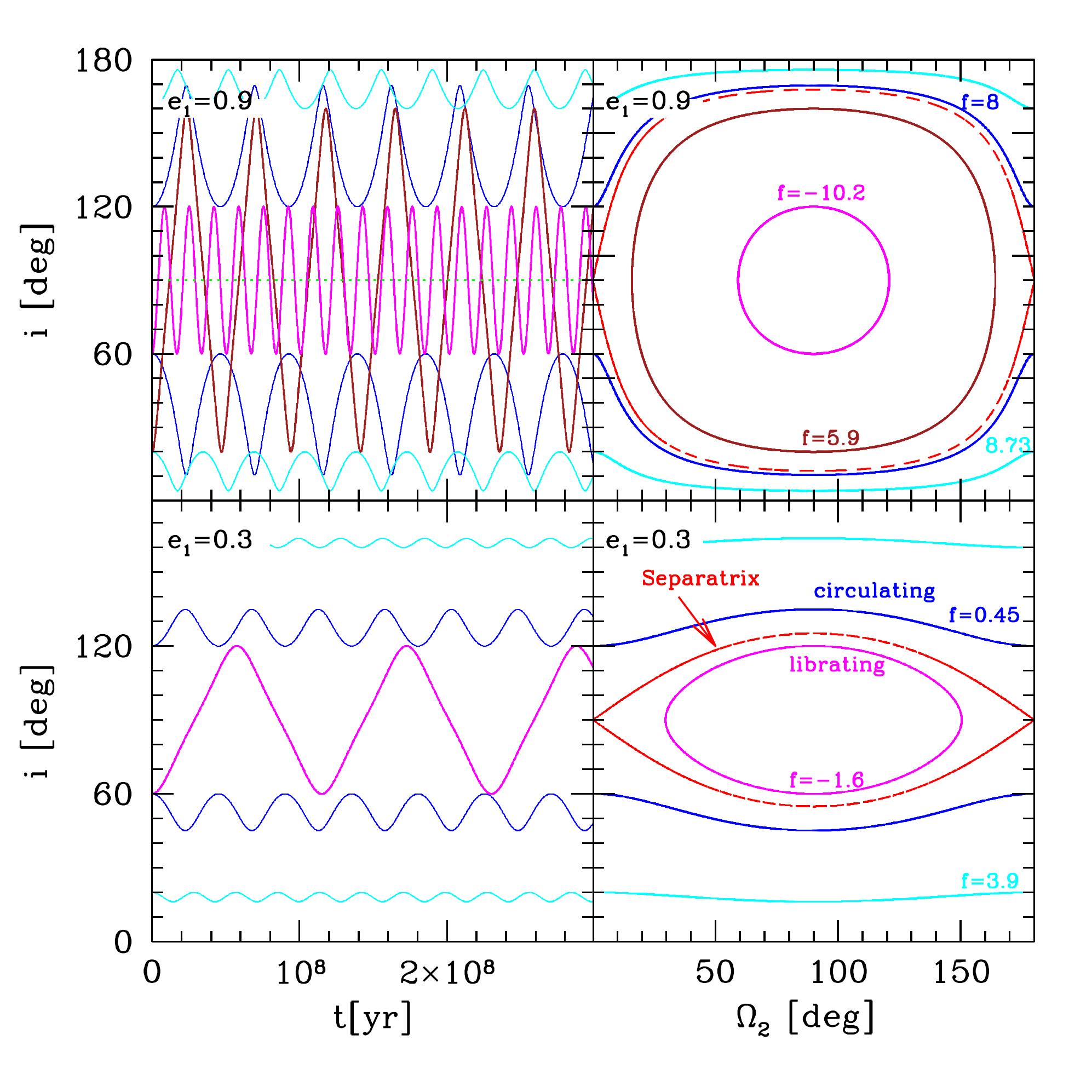}
\caption{\upshape {\bf The quadrupole-level of approximation evolution}.  We show the time evolution of the inclination in the left panels and the cross section trajectory in the inclination-$\Omega_2$ plane, in the right panels. We consider two cases, the top panels are for $e_1=0.9$ and the bottom panels are for $e_1=0.3$. We show the following examples: in the circulating mode (setting initially $\Omega_2=0$): $i=20^\circ$ (cyan), $60^\circ$ (blue), and in the librating mode (setting initially $\Omega_2=90^\circ$): $i=20^\circ$ (brown) and $60^\circ$ (magenta). The separatrix corresponds to $i=90^\circ$ for the two different inner eccentricities is shown in red. Note that in the case of $e_1=0.3$ there is no librating mode for $i=20^\circ$. 
%(as the minimum inclination of the separatrix is larger than $20^\circ$. 
For consistency we adopt the following orbital parameters: $m_1=1$~M$_\odot$, $m_1=1$~M$_J$, $a_1= 3$~AU and $a_2=40$~AU, $\omega_2=90^\circ$ and $e_2=0$.
} \label{fig:quad}\vspace{0.2cm}
  \end{center}
\end{figure}

The quadrupole-level of approximation is integrable and thus provides a good starting point.  
%The quadrupole-level of approximation describes the known behavior of precession of the nodes. 
Unlike the quadrupole-level approximation for the inner orbit test particle, here the z-component of the particle's  angular momentum is {\it not} conserved, as the Hamiltonian depends on $\Omega_2$. However, at this level 
$J_2$ is conserved, and thus the outer orbit eccentricity $e_2$ remains constant as the Hamiltonian does not depends on 
 $\omega_2$. The equation of motion for the inclination takes a simple form:
 \begin{eqnarray}\label{eq:dthetadtquad}
 %\frac{d\theta}{d t} = - \frac{15 a_1^2 k^4 m_1 m_2 e_1}{8 a_2^2 \sqrt{(a_2 k^2 (m_1+m_2))^3} } \frac{1-\theta^2}{(1-e_2^2)^2} \sin 2\Omega_2 \ ,
 \frac{d\theta}{d t}\bigg|_{\rm quad} & =&  \\&&  - \frac{15}{8} \left( \frac{a_1}{a_2}\right)^2  \frac{2\pi}{P_2} \frac{m_1m_2}{(m_1+m_2)^2} e_1^2  \frac{1-\theta^2}{(1-e_2^2)^2} \sin 2\Omega_2 \ ,\nonumber
 \end{eqnarray}
 where we consider the time evolution and not the scaled evolution for completeness, {and $P_2$ is the period of the outer orbit}.  We can find the maximum and minimum inclination by setting $\dot{\theta}=0$. % which reduces to:
%   \begin{equation}
%(1-\theta^2 )\sin 2\Omega_2  =0 \ . 
% \end{equation}
 %In other words, 
 Thus, we find  that the values of the longitude of ascending nodes that satisfy this condition   are $\Omega_2=n\pi/2$, where $n=0,1,2..$.  In other words, $\Omega_2$ has two classes of trajectories, librating and circulating. The trajectories in the librating region are bound between two values of $\Omega_2$, while the circulation region represents trajectories where the angles are not constrained between two specific values.
 On circulating trajectories, at $\Omega_2=0$, the inclination ($i<90^\circ$) is largest (where the $i>90^\circ$ case is a mirror image of the $i<90^\circ$ one). The extrema points  for the librating mode are located at $\Omega_2=90^\circ$. 
 The time evolution of $\Omega_2$ for the quadrupole-level of approximation is given by: 
  \begin{eqnarray}\label{eq:Omega2dot}
\frac{d\Omega_2}{dt}\bigg|_{\rm quad} &=& \\ & & -\frac{m_1 m_2}{(m_1+m_2)^2}\frac{2\pi}{P_2 }   \left( \frac{a_1}{a_2} \right)^2 \frac{3 \theta \left( 2+3e_1^2 -5 e_1^2 \cos 2\Omega_2 \right) }{8 (1-e_2^2)^2} \ . \nonumber
\end{eqnarray}
 In Figure \ref{fig:quad} we show the evolution associated for the quadrupole-level of approximation. The two  librating and circulating trajectories are considered, where we folded the $\Omega_2$ angle to be between $0-180^\circ$.  The librating mode gives the nominal precession of the nodes, at which  the inclination oscillates between the $i_{90}$ inclination (the inclination for which $\Omega_2=90^\circ$) and $180^\circ-i_{90}$. The precession of the nodes was noted before in the literature \citep[e.g.,][]{Inn+97}.
 
From the latter Equation and Equation (\ref{eq:dthetadtquad})  we have
   \begin{equation}\label{eq:OmTh}
\frac{d\Omega_2}{d\theta}=\theta \frac{ 2+3e_1^2-5e_1^2\cos 2\Omega_2}{5 e_1^2 (1-\theta^2)\sin 2 \Omega_2}  \ . 
 \end{equation}
 Integrating the two sides we have:
    \begin{equation}\label{eq:OmThInt}
\int^{\Omega_a}_{\Omega_b}\frac{\sin 2 \Omega_2   }{ 2+3e_1^2-5e_1^2\cos 2\Omega_2}  {d\Omega_2}=\int_{\theta_a}^{\theta_b}  \frac{\theta}{ 5e_1^2 (1-\theta^2) }  {d\theta} \ ,
 \end{equation}
 where $\Omega_{a,b}$ is the longitude of ascending nodes that is associated with the inclination value of $\theta_{a,b}=\cos i_{a,b}$.
 
  \begin{figure*}
\hspace{-2cm}
\centering
\begin{minipage}[b]{.4\textwidth}
\centering
\includegraphics[scale=.4]{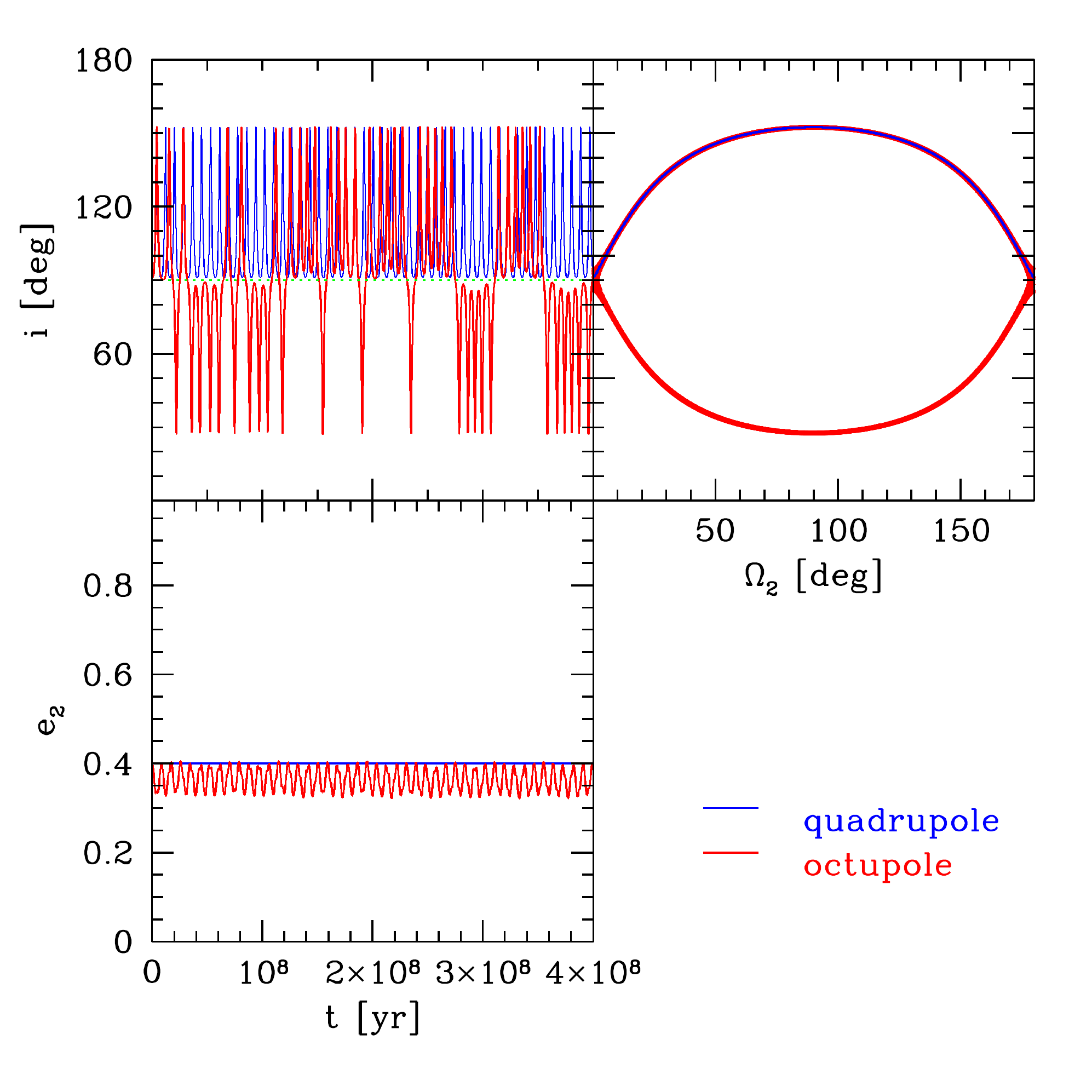}
\end{minipage}\qquad
\begin{minipage}[b]{.4\textwidth}
\centering
\includegraphics[scale=.4]{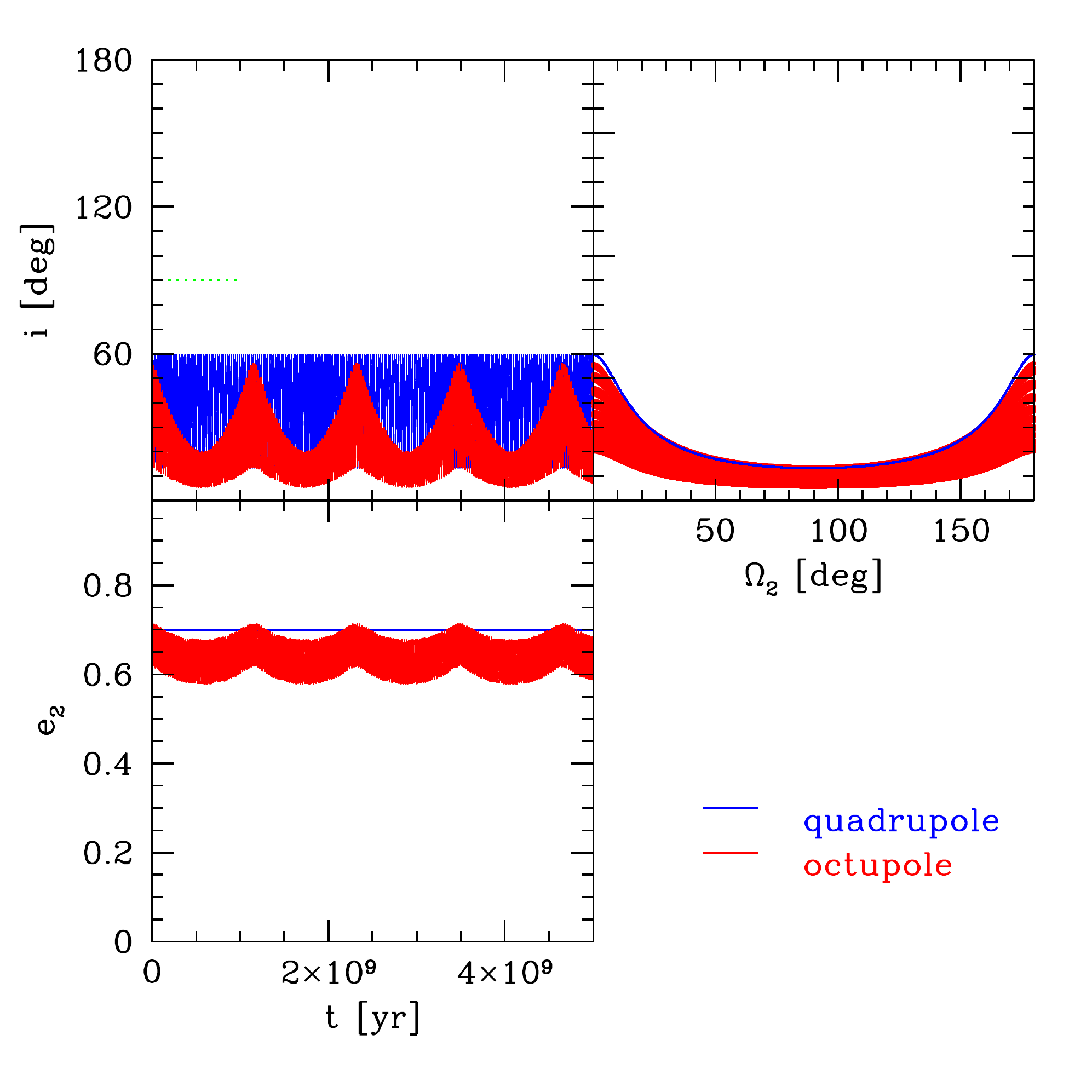}
\end{minipage}
\qquad\caption{ \upshape {\bf The role of octupole} We consider two cases, the secular evolution up to the quadrupole-level of approximation (blue lines) and up to the octupole level of approximation (red lines).  We show the time evolution of the inclination and the outer orbital eccentricity (which remains constant at the quadrupole-level of approximation). We also consider the inclination evolution as a function of $\Omega_2$.   {\bf Left panels:} We consider the following system:   $m_1=1$~M$_\odot$, $m_2=1$~M$_j$, $a_1=0.4$~AU, $a_2=7$~AU, $e_1=0.65$ and $e_2=0.4$. We initialize the system with $\omega_2=\Omega_2=0^\circ$ and $i=91^\circ$.  {\bf Right panels:}  $m_1=1$~M$_\odot$, $m_2=1$~M$_j$, $a_1=3$~AU, $a_2=50$~AU, $e_1=0.85$ and $e_2=0.7$. We initialize the system with $\omega_2=0^\circ$, $\Omega_2=40^\circ$ and $i=20^\circ$. %1.     0.001    0.4    7.   0.65   0.4    0.      0.         91.          400.
%1.     0.001    3.    50.   0.85   0.7    0.      40.         20.          9000.
}
\label{fig:quadoct}
\end{figure*}
 
For the circulating mode, we find that, after integrating over Equation (\ref{eq:OmTh}) from $i_{\rm max}$ to $i_{\rm min}$ (and $\Omega_2=0$ to $\Omega_2=90^\circ$, respectively) we get  
    \begin{equation}\label{eq:imaxmin}
\sin i_{\rm min}=\sin i_{\rm max} \sqrt{\frac{1-e_1^2}{1+4e_1^2} } \ . 
 \end{equation}
 Note that this expression can be also achieved by considering the conservation of energy between the minimum and maximum cases.
 Setting the initial conditions for the energy, Equation (\ref{eq:fquad}), we can find the extrema points as a function of the initial conditions.  
 A special case can be considered when $\Omega_2$ is initially set to be zero, and then the maximum inclination is the initial inclination $i_0$, in other words:     \begin{equation}\label{eq:imim} \sin i_{\rm min}=\sin i_{0} \sqrt{\frac{(1-e_1^2)}{(1+4e_1^2)}}  \quad  {\rm for } \quad \Omega_{2,0}=0^\circ \ .\end{equation} In other words we can set $i_{\rm max}\to 90^\circ$ and for a given $e_1$ find the largest $i_{\rm min}$ allowed, which corresponds to the separatrix. This relationship  is also apparent in figure 14 in \citet{Zanardi+17} numerical results. As can be seen, for the circulating mode depicted in Figure \ref{fig:quad}, setting initially $i=20^\circ$ corresponds to $i_{\rm min}  = 16.25^\circ$ and $10.56^\circ$ for $e_1=0.3$ and $e_1=0.9$, respectively, consistent with Equation (\ref{eq:imim}). A comparison between the flip criterion and the numerical results can be seen in Figure \ref{fig:TwoPanle_e5}. 
 
 For the librating mode we set $i_{\rm min}$ which associated with the $\Omega_2=90^\circ$ case, and thus integrating over Equation (\ref{eq:OmTh}) between $i=90^\circ$ to $i_{\rm min}$ (and $\Omega_2=\Omega_{2,{\rm min}}$ to $\Omega_2=90^\circ$, respectively), we get,
     \begin{equation}\label{eq:Omegamin}
\cos 2\Omega_{2,\rm min} = \frac{2+3e_1^2-(2+8e_1^2)\sin^2 i_{\rm min}}{5e_1^2} \ . 
 \end{equation}
 Thus, setting initially $\Omega_2=90^\circ$ as in the examples depicted in Figure \ref{fig:quad} we find that the minimum value that $\Omega_2$ can achieve in the $e_1=0.9$ case is $59.23^\circ$ for the  $i_{\rm min}=60^\circ$ and $15.95^\circ$ for the $i_{\rm min}=20^\circ$ example, consistent with the numerical results. 
Hence, the range of which $\Omega_2$ is librating on is $2\times (90^\circ-\Omega_{2,{\rm min}})$

 \subsection{Timescale}\label{sec:Time}
The timescale associated with the evolution can be  estimated  from the equation of motion for $\Omega_2$ at the quadrupole-level [Equation (\ref{eq:Omega2dot})], 
%For the liberating mode the maximum and minimum inclination do not depend on $e_1$. 
%(3 a1^2 k2^2 m1 m2 \[Theta] (-2 - 3 e1^2 + 
%   5 e1^2 Cos[2 \[CapitalOmega]2]))/(8 a2^2 (-1 + 
  % e2^2)^2 (a2 k2 (m1 + m2))^(3/2))
%\begin{eqnarray}
%\frac{d\Omega_2}{dt}\bigg|_{\rm quad} &=& \\ & & \frac{m_1 m_2}{P_2 (m_1+m_2)^2}   \left( \frac{a_1}{a_2} \right)^2 \frac{3 \theta \left( 2+3e_1^2 -5 e_1^2 \cos 2\Omega_2 \right) }{8 (1-e_2^2)^2} \ , \nonumber
%\end{eqnarray}
%where we have used the actual time here, using Equation (\ref{eq:time}).  
 for the circulating mode, by  
setting $d \Omega_2\to \pi$ and taking the terms in the parenthesis to be roughly order of unity (which is achieved, by setting $\Omega_2\to 0$):
\begin{equation}
t_{\rm  quad}  \sim \frac{4}{3} P_2 (1-e_2^2)^2 \frac{ (m_1+m_2)^2}{m_1 m_2}  \left( \frac{a_2}{a_1} \right)^2  \quad {\rm circulating.} %\frac{(1-e_2^2)^2}{\cos i}  \ .
\end{equation}
For the example system depicted  in Figure \ref{fig:quad}, this equation gives a timescale of about $6\times 10^7$~yr, for initial inclination of $60^\circ$, which agrees with the circulating mode (although we note that different $e_1$ give slightly different timescales).  We also estimate the timescale in the librating mode by setting $d \Omega_2\to 2\times (90^\circ-\Omega_{2,{\rm min}})$, which we have found earlier. Thus,
\begin{eqnarray}
t_{\rm  quad}  &\sim & 2\times \frac{ 2 (\pi/2-\Omega_{2,{\rm min}})}{2\pi} \frac{8}{3} P_2 (1-e_2^2)^2 \times \nonumber \\ && \frac{ (m_1+m_2)^2}{m_1 m_2}  \left( \frac{a_2}{a_1} \right)^2  \quad {\rm librating.} %\frac{(1-e_2^2)^2}{\cos i}  \ .
\end{eqnarray}
The $2$ pre-factor of here comes from numerical comparisons  to the examples depicted in Figure \ref{fig:quad}. Note the $e_1$ dependency  that  rises from    $\Omega_{2,{\rm min}}$. 
This timescale is consistent with % in good agreement with 
the examples depicted in Figure \ref{fig:quad} by less than a factor of two.

\section{The role of the octupole level of approximation }\label{sec:oct}

The octupole level of approximation can significantly affect the overall dynamics of the general hierarchical three body system \citep[see][and reference therein]{Naoz16}.
Specifically, in the {\it inner} test particle case, the inner orbit's z component of the angular momentum  is not conserved anymore and the orbit is allowed to flip \citep[for large range of initial inclinations][]{LN,Li+13,Li+14Chaos}. In our case, the $J_{2,z}$ is not conserved at the quadrupole-level, but $J_2$ is. Thus, the octupole level of approximation in this case allows for variations of $e_2$ and introduces higher level resonances, which may result in a chaotic behavior (see below). 

\begin{figure}
\begin{center}
\includegraphics[width=\linewidth]{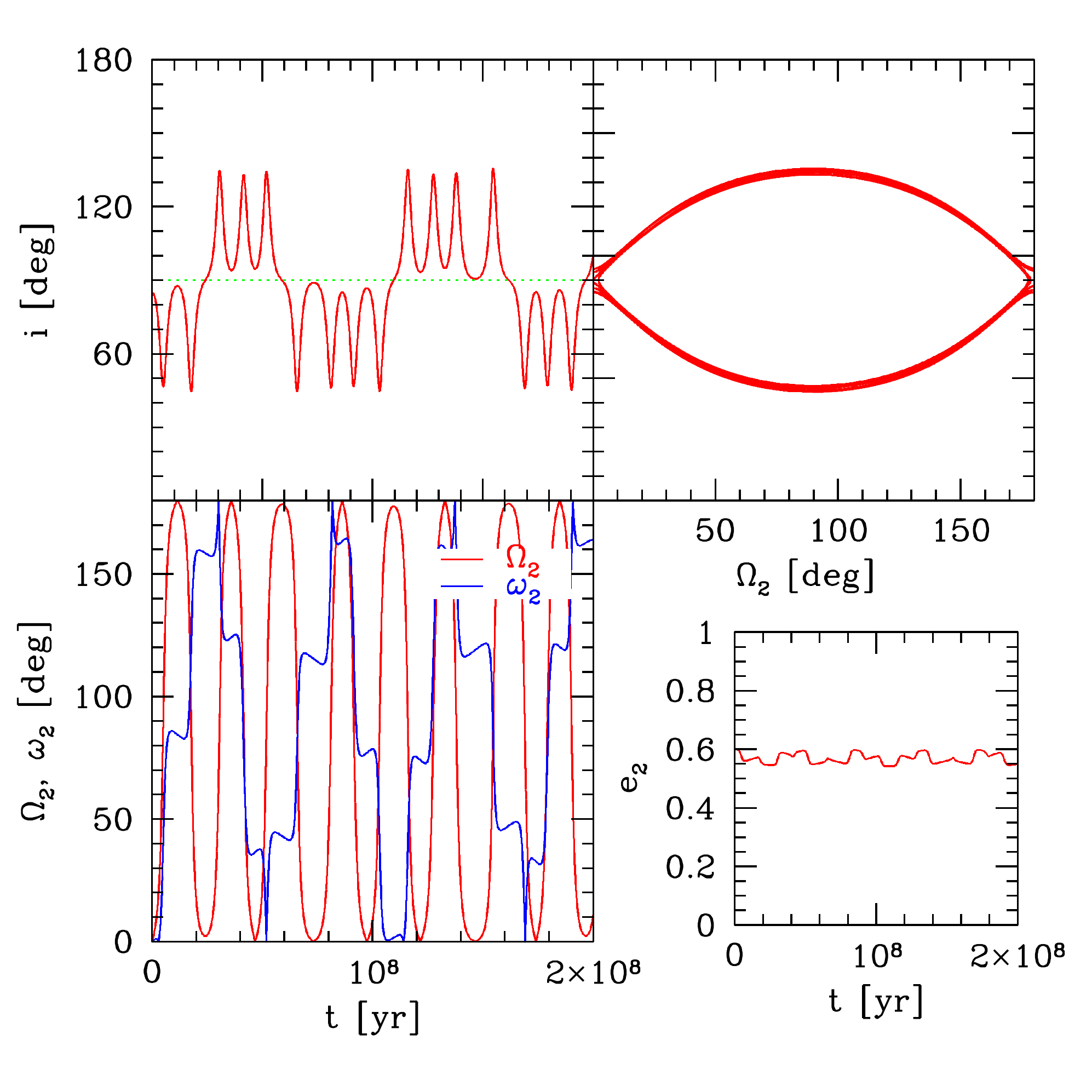}
\caption{\upshape We consider the system: $m_1=1$~M$_\odot$, $m_2=1$~M$_j$, $a_1=0.5$~AU, $a_2=10$~AU, $e_1=0.4$ and $e_2=0.6$. We initialize the system with $\omega_2=\Omega_2=0^\circ$ and $i=85^\circ$.  We show the time evolution of the orbital parameters, i.e., argument of pericenter $\omega_2$, longitude of ascending node $\Omega_2$, inclination $i$  and the outer orbital eccentricity, $e_2$. We also consider the inclination evolution as a function of $\Omega_2$.  Here both $\Omega_2$ and $\omega_2$ were folded to achieve the $0-180^\circ$ symmetry.     %{\bf 1.     0.001    0.5    10.  0.4  0.6    0.      0.         85.          100. }. 
} \label{fig:Ex1} 
  \end{center}
\end{figure}

In Figure \ref{fig:quadoct} we consider two representative example for which we compare the quadrupole (blue lines)  and octupole (red lines) levels of approximation, where we consider the time evolution of the eccentricity and inclination. In both of these examples we consider a $1$~M$_\odot$ star orbited by an eccentric Jupiter, with a test particle on a far away orbit. One can consider such a setting to represent a result of a scattering event for example. 

On the left set of panels of Figure \ref{fig:quadoct} we consider a Jupiter at $0.4$~AU with $e_1=0.65$ and a test particle at $7$~AU with $e_2=0.4$, initialized on a retrograde orbit ($i=91^\circ$). With the introduction of the octupole level of approximation to the calculation, the test particle eccentricity starts to oscillate, though in this case it never increases pass its initial value (due to choice of initial conditions here). More notably, the test particle inclination, with respect to the total angular momentum, oscillates from retrograde ($>90^\circ$ which was the initial condition) to prograde ($<90^\circ$). As in the more general case, there is no apparent associated timescale for this flipping modulation and it seems chaotic in nature (see below). While the quadrupole-level is circulatory in nature (see $i-\Omega_2$ plot) a libration behavior emerges at the octupole level. 

On the right set of panels of  Figure \ref{fig:quadoct} we consider a Jupiter at $3$~AU with $e_1=0.85$ and a test particle at $50$~AU with $e_2=0.7$, and the system is initialized on a prograde orbit ($i=20^\circ$). Here, like the previous example the outer test particle eccentricity begins to oscillate, and even grows above the initial value. However, unlike the   previous example this system does not  flip. The inclination does oscillate with a long scale modulation and we show the long scale evolution that captures about four octupole cycles. The system does not exhibit a chaotic behavior in this case, and it remains in a circulatory trajectory even after the inclusion of the octupole level of approximation to the calculation.

In Figure \ref{fig:Ex1} we zoom-in on the evolution of a different example, and also provide the time evolution of $\omega_2$ for the octupole level itself. As in the general hierarchical secular three body problem we find the short scale (associated with the quadrupole) oscillations, that are modulated by the higher level octupole approximation. 
The octupole modulations take place on timescales which is between few$\times t_{\rm quad}$ to few tens$\times t_{\rm quad}$. 

\begin{figure*}
\begin{center}
\includegraphics[width=\linewidth]{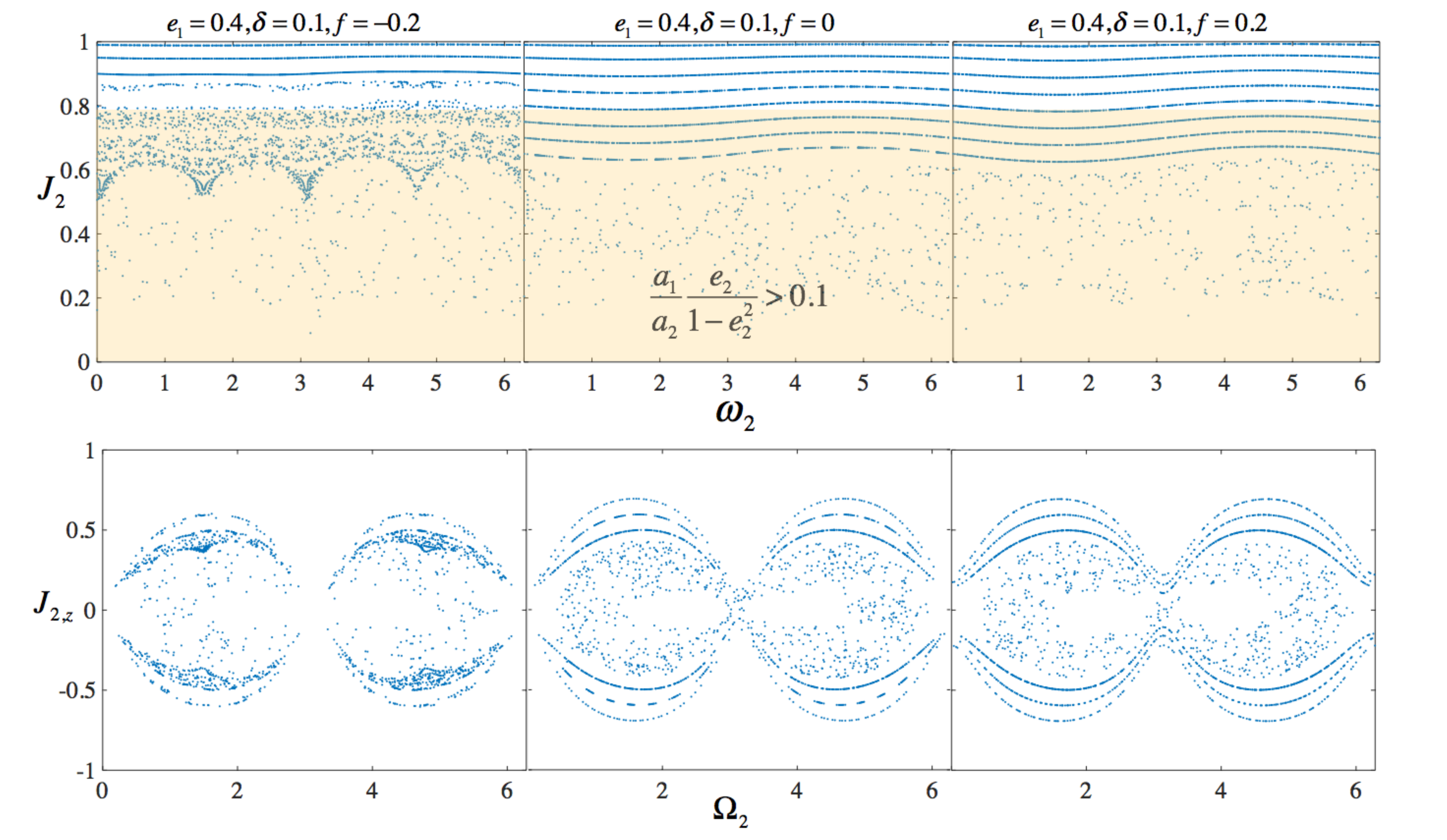}
\caption{\upshape {\bf Surface of section}. We consider $e_1=0.4$, $\delta=0.1$ and, from left to right $f=-0.2,0$ and $0.2$. We note that the cases $f=10$ and $f=-10$ gives a similar map to the $f=0.2$ and $f=-0.2$ respectively  and thus, were not depicted, to avoid clutter.   The light orange stripe marks {the unstable} regime, for which $\epsilon>0.1$.  } \label{fig:sur1}
  \end{center}
\end{figure*}

Its interesting to note that the inclination flips shown in Figures \ref{fig:quadoct} and \ref{fig:Ex1} are {\it qualitatively} different from the ordered, back and forth oscillation of the quadrupole-level of approximate evolution.  The latter produces a   simple, ordered oscillation of the inclination angle between $i_0$ and   $180^\circ- i_0$, for the librating regime. However, in the presence of the octupole-level of approximation the system behaves similarly to the general flips discussed in \citet{Naoz11}, where the inclination oscillates for sometime at the  prograde ($i<90^\circ$) regime, and then flips to retrograde configuration ($i>90^\circ$).

The eccentricity, $e_2$, gives rise to an additional complication as in essence the outer test particle eccentricity can grow so much until the orbits will cross. We adopt the nominal stability criterion 
\begin{equation}
\epsilon = \frac{a_1}{a_2}\frac{e_2}{1-e_2^2} < 0.1
\end{equation}
to guid us when the system leaves stability. %\citet{Naoz+12GR} showed that this carrion has 
We discuss this stability criterion in the context of N-body comparisons in appendix \ref{sec:Nbody}.

\begin{figure*}
\begin{center}
\includegraphics[width=\linewidth]{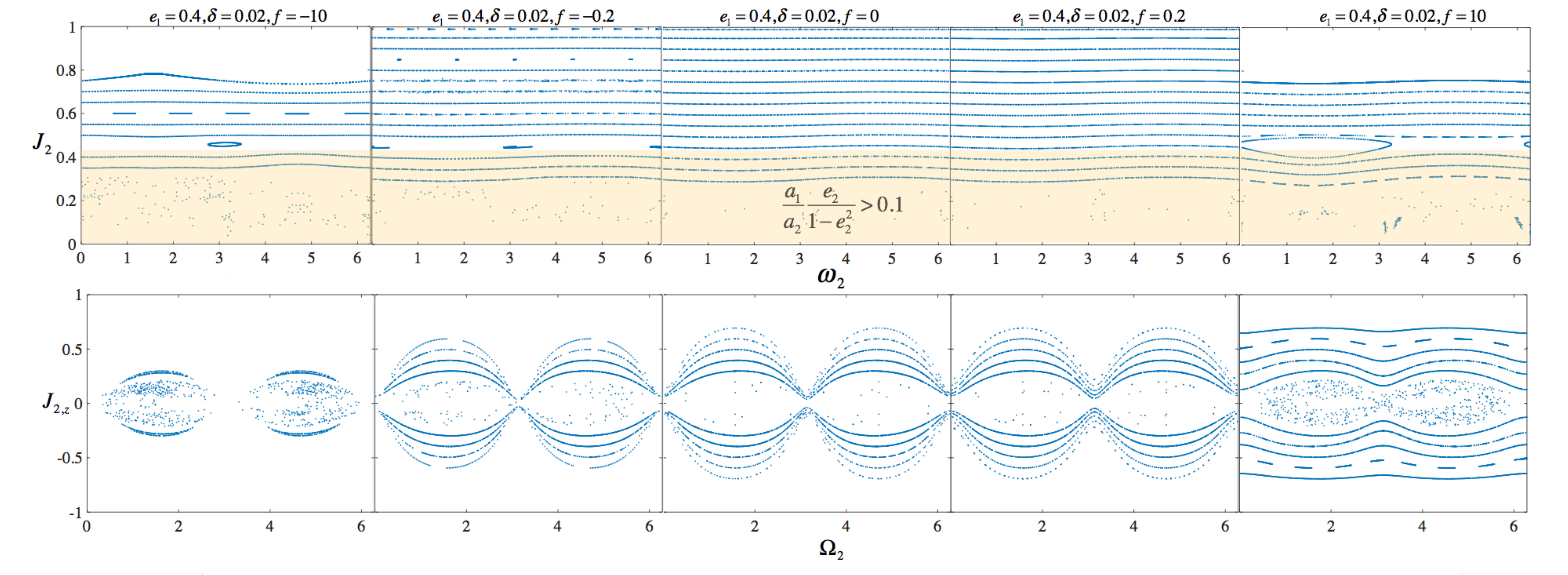}
\caption{\upshape {\bf Surface of section}. We consider $e_1=0.4$, $\delta=0.02$ (compared to Figure \ref{fig:sur1}, this means changing the factor $(m_1-m_2) a_1/a_2 /(m_1+m_2)$ by factor of $5$) and, from left to right $f=-10,-0.2,0,0.2$ and $10$. The light orange stripe marks {the unstable regime}, for which $\epsilon>0.1$.  } \label{fig:sur2}
  \end{center}
\end{figure*}

\begin{figure*}
\begin{center}
\includegraphics[width=\linewidth]{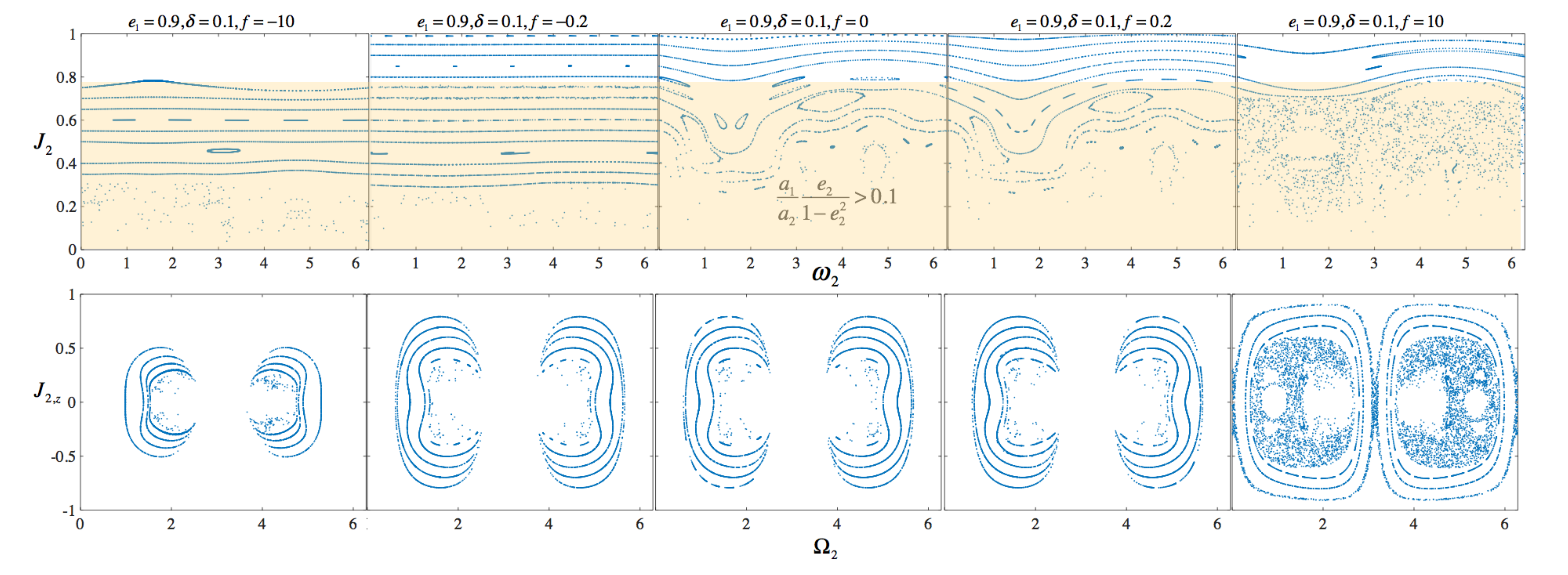}
\caption{\upshape {\bf Surface of section} We consider $e_1=0.9$, $\delta=0.1$  and, from left to right $f=-10,-0.2,0,0.2$ and $10$. The light orange stripe marks {the unstable} regime, for which $\epsilon>0.1$. } \label{fig:sur3}
  \end{center}
\end{figure*}

%\section{Surface of Sections}
%{\bf Gongjie, please read carefully below and revised if needed}:\\
%{\bf the Jz-O plots are not updated. I just forwarded you the updated plots requiring $\omega$ decreases.}\\
To explore the chaotic nature of the system and the different dynamical regimes we use surface of sections. 
The outer test particle  approximation reduces the general hierarchical three body system from six degrees of freedom to four degrees of freedom. In addition, in the test particle limit, the inner orbit is stationary and reduces the system to two degrees of freedom. In this system $f$ and $\delta$ are the only conserved parameters and $\omega_2$ and $\Omega_2$ are the only coordinates that can change with time. For a two-degrees-of-freedom system, the surface of section projects a four-dimensional trajectory on a two-dimensional surface, where we select intersections of the trajectories on the surfaces when $\omega_2$ and $\Omega_2$ moves in the positive directions. For simplicity, we separate the two initial conditions into three characteristic parameters: $e_1$, the inner orbit eccentricity, which remains constant during the evolution, the energy, or initial value of the reduced Hamiltonian $f$, and \begin{equation*} \delta=\frac{m_1-m_2}{m_1+m_2}\frac{a_1}{a_2} e_1 \ . \end{equation*} Note that the energy depends on $\delta$ and $e_1$, so in fact, although we choose to characterize the surface of sections by three parameters, there are only two independent ones. 
%We identify three 
 %distinct regions in the surface of section: ``resonant regions,"``circulation regions," and ``chaotic regions".

In Figures  \ref{fig:sur1}-\ref{fig:sur3} we consider the surface of sections for various values of the $f$, $e_1$ and $\delta$, in the $J_2-\omega_2$ plane (top row in each figure) and in $J_{2,z}-\Omega_2$ plane (bottom row in each figure).
%{\bf Gongjie, can you please add here?}
% The latter plane is defined by setting $\omega_2=0$ and thus represents a manifold of the two possible solutions (the $i<90^\circ$ and the $i>90^\circ$).
In both planes we identify the  resonances at  which the momenta and angles undergo bound oscillations. The trajectories in this region are quasi-periodic, and the system is in the libration mode.  
The circulation region represents trajectories where the angles are not constrained between two specific values. Both { librating} and circulatory trajectories are mapped onto a one-dimensional manifold on the surface of section and they form lines on the section. However, chaotic trajectories are mapped onto a  higher dimensional % (since it may be fractional)
manifold and they are filling an area on the surface. We note that in some of the trajectories in the Figures, due to sampling limitation, seem as dashed lines, but they actually represent a one-dimensional manifold. In all of the maps we indicate the instability regime (light orange stripe) for which $\epsilon>0.1$.

{We intersect the trajectories at $\Omega_2 = \pi/2$ to produce the surfaces in the $J_2-\omega_2$ plane, in order to capture the librating cases. The empty regions at large $J_2$ at the parameter spaces in the far left and right panels in Figure \ref{fig:sur2} (i.e,. for the parameters: $e = 0.4$, $\delta = 0.02$ and $f = \pm 10$) and the far left panel in  Figure \ref{fig:sur3} ( for the parameters: $e = 0.9$, $\delta = 0.1$ and $f = -10$)  correspond to regions with no physical solutions. The variabilities in $J_2$ are mostly small in the stable regime. We see that there are regular behavior, i.e., trajectories which {fill one-dimensional lines on the surface of section} in most of the stable region.} We find the emergence of chaos in  {parts of the unstable zones, in particular when $J_2$ is low ($e_2$ is high).} 

Considering the $J_{2,z}-\Omega_2$ plane, {we intersect the trajectories at $\omega_2 = 0$.} The system exhibits a chaotic behavior across all parameter regime of $e_1,\delta$ and $f$. Most of the circulation region, associated with curve, non chaotic one-dimensional manifold, are typically associated with $|J_{2,z}|\gsim 0.3$. {The outer orbits can flip ($J_{2, z}$ shifts signs) in most of the parameter space.}

{Resonances can be easily identified in a few of the maps. Specifically, in the $J_2-\omega_2$ plots, the resonances can be found centered near $\bf \omega_2 = \pi$ (e.g., $e_1 = 0.4$, $\delta = 0.02$, $f = -10$, and $e_1 = 0.9$, $\delta = 0.1$, $f = -10$, {etc.}), {and $\omega_2 = \pi/2$ (e.g., $e_1 = 0.4$, $\delta = 0.02$, $f = 10$)}. The dynamics is quite complicated when $e_1$ is higher and when $\delta$ is larger, and higher order resonances (appearing as small liberating islands) emerge in the surface of section in the $J_2-\omega_2$ plane when $e_1 = 0.9$, $\delta = 0.1$ and $f=0$. 
Resonances can also be identified in the $J_{2,z}-\Omega_2$ plane, such as $e_1=0.9$, $\delta=0.1$, $f=10$.
%Resonances
%can be identified in the $J_{z,2}-\Omega$ plane e.g., $e_1=0.9$, $\delta=0.1$, $f=10$. 
}
%In other words, while quasi-periodic trajectories form lines on the section, chaotic trajectories are area-filling. Embedded in the chaotic region, the small islands correspond to the higher order resonances, which are caused by the interaction between the primary resonances. The trajectories in the higher order resonant regions are also quasi-periodic.

\section{The Role of the General Relativity}\label{sec:GR}

\begin{figure}
\begin{center}
\includegraphics[width=\linewidth]{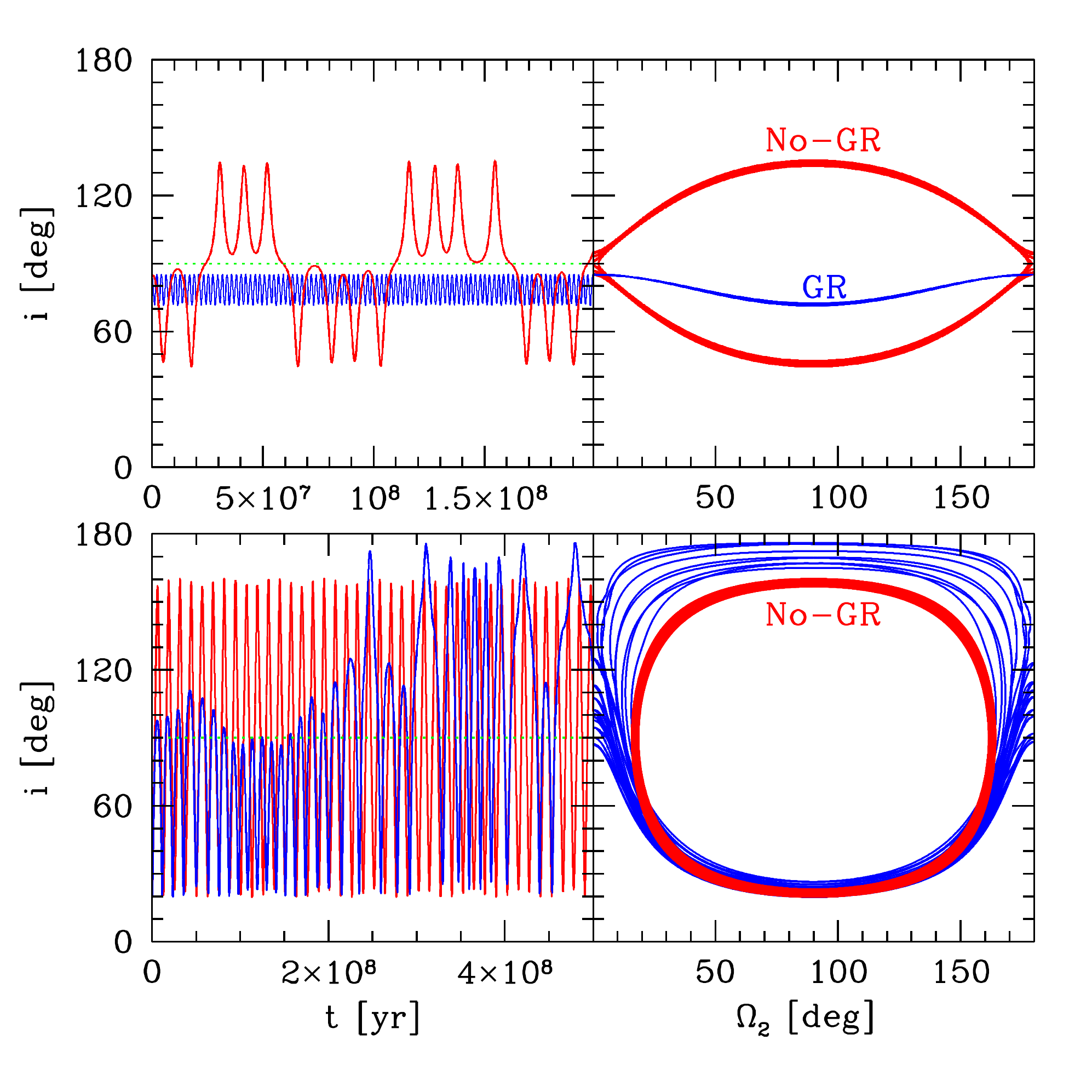}
\caption{\upshape {\bf The role of GR.} We consider two cases, the evolution without GR (red lines) and the evolution with GR (blue lines). {\bf  Top panel}:  We consider the following system: $m_1=1$~M$_\odot$, $m_2=1$~M$_j$, $a_1=0.5$~AU, $a_2=10$~AU, $e_1=0.4$ and $e_2=0.6$. We initialize the system with $\omega_2=\Omega_2=0^\circ$ and $i=85^\circ$. For this system we find that  $t_{\rm GR,inner}\sim 5\times 10^6$~yr which is much shorter than the quadrupole timescale.  {\bf Bottom panel}:  We consider the following system: $m_1=1$~M$_\odot$, $m_2=1$~M$_j$, $a_1=3$~AU, $a_2=40$~AU, $e_1=0.9$ and $e_2=0.65$. We initialize the system with $\omega_2=90^\circ$ and $\Omega_2=100^\circ$ and $i=20^\circ$. The GR precession timescale for this system is estimated as $t_{\rm GR,inner}\sim 4.4\times 10^8$~yr, which is longer than the quadrupole timescale. This is a typical situation to an individual debris disk particle  or a  icy body reservoir  object (see section \ref{sec:Kuip}). \label{fig:GR}}
  \end{center}
\end{figure}

As noted previously in many studies, General Relativity (GR) precession tends to suppress the inner orbit eccentricity excitations associated with the {Eccentric Kozai-Lidov} (EKL) mechanism, and thus suppress the flips \citep[e.g.,][]{Naoz+12GR}. In our secular case, the inner orbit is massive and the outer orbit is a test particle, so practically the inner orbit does not feel the outer orbit gravitational interactions. However, the inner orbit can still precess due to GR with the nominal precession rate \citep[e.g.,][]{Naoz+12GR}:
\begin{equation}\label{eq:GR1}
\frac{d\omega_1}{dt} \bigg |_{\rm GR,inner} = \frac{3 k^{3} (m_1 + m_2)^{3/2} }{a_1^{5/2} c^2 (1 - e_1^2)} \ ,
\end{equation}
{where $k^2$ is the gravitational constant and $c$ is the speed of light. }
However, in our frame of reference, where the inner orbit carries all the angular momentum, we are basically working in the rotating frame of the inner orbit. Therefore, since we set $\omega_1=-\pi-\Omega_2$, (see Appendix \ref{App:Ham}), GR precession of $\omega_1$  translates to a precession of $\Omega_2$. Thus, using our coordinate transformation we find: 
 %of the inner  we use our coordinate transformation,  $\omega_1=-\pi-\Omega_2$ (see Appendix \ref{App:Ham})  and find:
\begin{equation}\label{eq:GR2}
\frac{d\Omega_2}{dt} \bigg |_{\rm GR,\Omega_2} = -\frac{3 k^{3} (m_1 + m_2)^{3/2} }{a_1^{5/2} c^2 (1 - e_1^2)} \ ,
\end{equation}
which can suppress the inclination oscillations. 
The timescale associated with that precession is the  nominal GR one:
\begin{equation}\label{eq:tGR}
t_{\rm GR,\Omega_2} \sim 2\pi \frac{a_1^{5/2} c^2 (1-e_1^2) }{ 3 k^3 (m_1+m_2)^{3/2}} \ .
\end{equation}
In Figure \ref{fig:GR} we consider two examples, one for which  $t_{\rm GR,\Omega_2}\sim 5\times 10^6$~yr is  smaller than $t_{\rm  quad}\sim 6\times 10^6$~yr (top panel), and the other of which $t_{\rm GR,\Omega_2}\sim 4.4\times 10^8$~yr is a bit longer than the corresponding  $t_{\rm  quad}\sim 2\times 10^7$~yr (bottom panel). 
%Specifically, the top panel of Figure \ref{fig:GR} we consider a system for which $t_{\rm GR,inner}\sim 5\times 10^6$~y is much smaller compared to the quadrupole timescale  
Both of these examples had a  Sun size star and a Jupiter planet, orbited by a far away test particle. The Jupiter has a non-negligible eccentricity, that perhaps can be a result of  either a scattering event or a high eccentricity migration.  In the top panel the Jupiter was set at $a_1=0.5$~AU with $e_1=0.4$ and the test particle was set at $a_2=10$~AU and $e_2=0.6$. The system in the absence of GR was in libration mode and as noted before exhibited a chaotic nature. However, the GR precession bound the system into a circulatory regime, and suppressed the flips.   In the bottom panel the Jupiter was set at $a_1=3$~AU and $e_1=0.9$, while the test particle was set at $a_2=40$~AU with $e_2=0.65$. Note that in this latter case, although the Jupiter is rather far from the host star and the GR precession timescale is longer than the secular precession timescale, the GR changes the dynamics. Specifically, before the inclusion of GR precession, the system was in a libration mode and seemed quasi-periodic, however, after the inclusion of GR the system exhibits both libration and circulation, and the emergence of chaotic behavior seems to take place.  The dramatic change in dynamical behavior with the inclusion of GR precession, even if takes place on {\it longer} timescales than the secular timescales, was noted previously  in the general and inner test particle case in \citet{Naoz+12GR}.

We note that in all of our calculation below we also take into account the outer orbit GR precession \citep[e.g.,][]{Naoz+12GR}:
\begin{equation}\label{eq:GR2}
\frac{d\omega_2}{dt} \bigg |_{\rm GR,outer} = \frac{3 k^{3} (m_1 + m_2)^{3/2} }{a_2^{5/2} c^2 (1 - e_2^2)} \ ,
\end{equation}
which typically takes place on much larger timescales.

\section{A study case application: individual Debris disk particles }\label{sec:Kuip}

% \subsection{Debris disks and Kuiper-belt like objects}\label{sec:Kuip}
\begin{figure}
\begin{center}
\includegraphics[width=\linewidth]{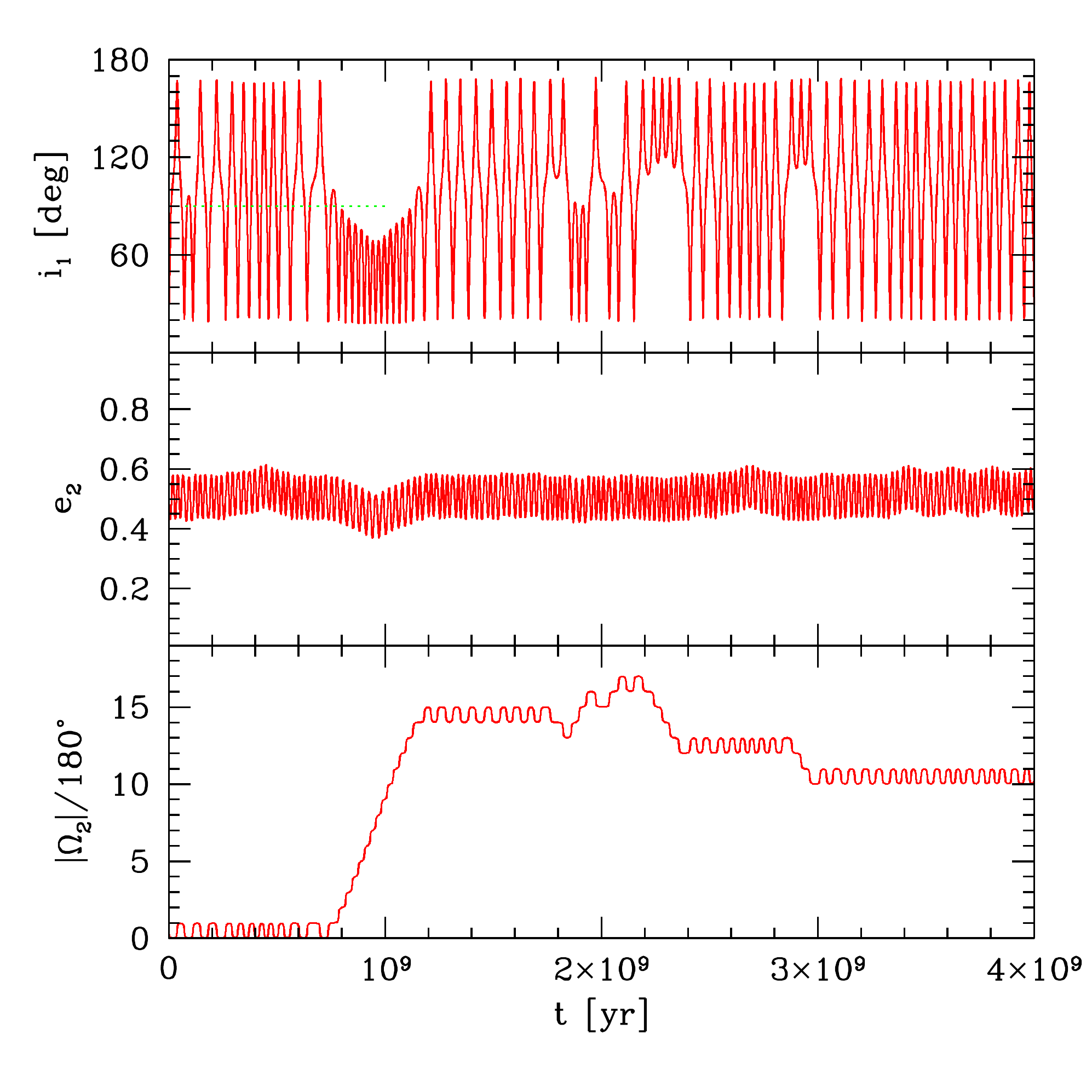}
\caption{\upshape {\bf Debris disk particles}. We consider the time evolution of a test particle located at $55$~AU from a $~$~M$_\odot$ star due to the gravitational perturbations from an eccentric Jupiter $a_1=5$~AU and $e_1=0.85$. We initialized the system with $e_2=0.5$, $\Omega_2=90^\circ$, $\omega_2=0^\circ$ and $i=20^\circ$. 
We consider from top to bottom the inclination, eccentricity and $\Omega_2$.   The transition between librating and circulating can clearly be seen in the bottom panel. When the angle is in circulation mode, it increases in value as a function of time.
% Thus, when $\Omega_2$ increases as a function of time, the system is in a  librating mode. 
  } \label{fig:Kuip1}
  \end{center}
\end{figure}

Debris disks mark the late end stages of planet formation, and are made of  the  leftover  material of rocks and ices.
The gravitational interactions between these particles and interior or exterior companion can leave a distinct imprint on the morphology of the disk and can cause dust production \citep[e.g.,][]{Rodigas2014,Matthews2014,Nesvold2015,Lee2016,Nesvold2016}. Many of these studies typically focus on few million years of integration to allow for comparison of observations which usually can detect young systems. Here we allow for longer integration timescales and investigate the evolution of a test particle under the influence of an eccentric Jupiter. 

\begin{figure*}
\begin{center}
\includegraphics[width=\linewidth]{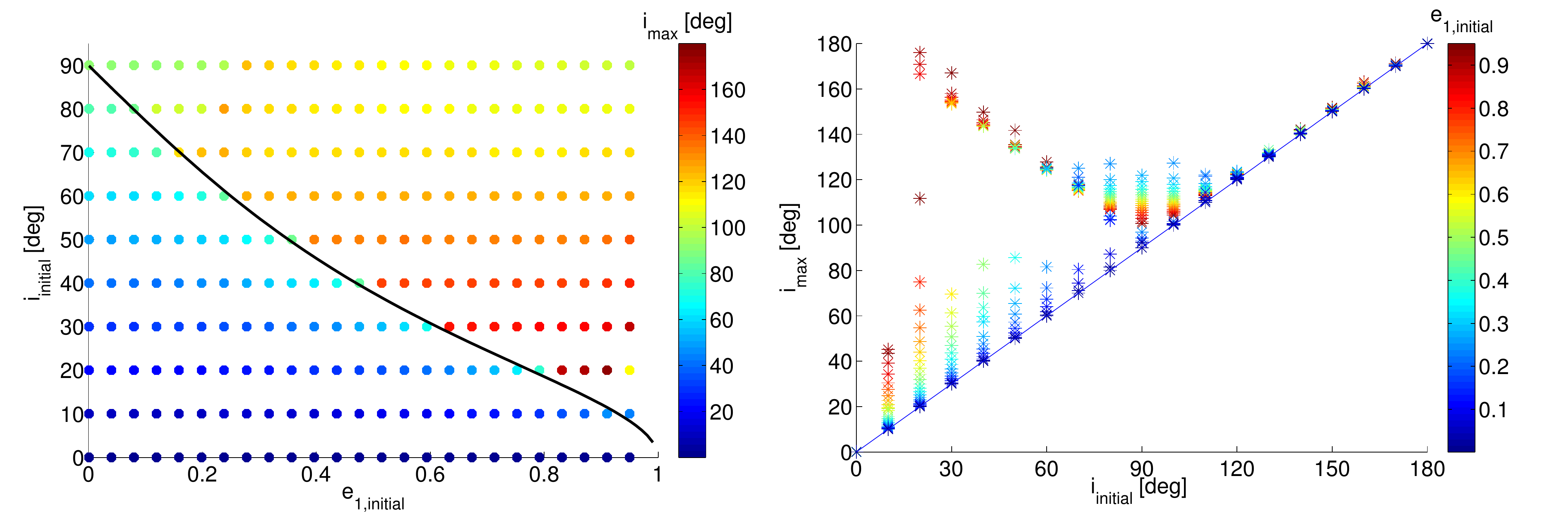}
\caption{\upshape {\bf Debris disk particles}. We consider a Sun-Jupiter like system, setting Jupiter at $5$~AU. We set the test particle at $55$~AU, with $e_2=0.5$. We vary systematically the inclination and Jupiter's eccentricity. We set the system initially with $\Omega_2=90^\circ$ and $\omega_2=0^\circ$. Left panel shows the  initial condition map (inclination vs the Jupiter's eccentricity) where the color code depicts the maximum inclination the system reached during its evolution of $4$~Gyr. The solid line shows the analytical equation to reach $90^\circ$, Equation  (\ref{eq:imaxmin}). The right panel shows the maximum inclination reached as a function of the initial inclination  (that goes all the way to $180^\circ$). The color code here marks the inner orbit's eccentricity. } \label{fig:TwoPanle_e5}
  \end{center}
  \vspace{0.3cm}
\end{figure*}

In Figure \ref{fig:Kuip1} we show an example system, where we consider an eccentric Jupiter at $5$~AU, with $0.85$ eccentricity orbiting a one solar mass star. The test particle is located at $55$~AU and initialized with an eccentricity of $0.5$.   We integrate the octupole level equations of motion in the presence of GR precession for both the inner and outer orbits. As in the example depicted in Figure \ref{fig:GR}, which also considered an icy body or comet reservoirs
%Kuiper-Belt /Tans Neptunian 
  analogs, the  orbit switches between 
 libration and circulation as can be seen in the bottom panel.
 % As the system in libration mode $\Omega_2$ winding up and thus, increasing in value. 

% 
% \begin{figure}
%\begin{center}
%%\includegraphics[width=\linewidth]{e_dis_obsNew}
%\includegraphics[width=\linewidth]{ExamplesCBKup}
%\caption{\upshape {\bf Astrophysical examples.} {\bf Top panel:} We consider the time evolution of a circumbinary planet around a stellar binary with $m_1=1$~M$_\odot$ and $m_2=0.5$~M$_\odot$ with $a_1=0.5$~AU and $e_1=0.75$ \citep[note that the inner orbital parameters are consistent with][simulations]{NF}. The planet is located $5$~AU away from the star and set initially with $i=85^\circ$ and $\Omega_2=\omega_2=0^\circ$.  {\bf Bottom panel:} We consider the evolution of a Kuiper-belt like object due to the gravitational perturbation of an inner eccentric Jupiter mass planet with $a_1=3$~AU and $e_1=0.9$ around a solar mass star. The test particle has $a_2=40$~AU and $e_2=0.65$ and initially set with $i=20^\circ$, $\omega_2=90^\circ$ and $\Omega_2=100^\circ$.  
%  } \label{fig:App1}
%  \end{center}
%\end{figure}
%

 We consider the effect of the planet's eccentricity, $e_1$, and the test particle inclination by surveying the parameter space of $e_1$ and initial inclination for a given system, where we set $a_2=55$~AU, with $e_2=0.5$, and set the system initially with $\Omega_2=90^\circ$ and $\omega_2=0^\circ$. Many giant exoplanets have high eccentricities, specifically for   giant planets ($m\sin i>0.1$~M$_J$, with separation $>0.05$~AU) the average eccentricity is $\sim 0.2$, and a maximum value of $0.97$\footnote{Taken from The Exoplanet Orbit Database \citep{Wright+11}.}. Thus, an eccentric Jupiter doesn't seem like an unlikely configuration for a planetary system. In the example depicted in Figure \ref{fig:Kuip1} the time evolution of the test particle's inclination  that starts with a moderate eccentricity oscillates between extreme values $20^\circ \sim 160^\circ$ without increasing its eccentricity.

  \begin{figure}
 \hspace{-0.8cm}
 \vspace{0.02cm}
 %\raggedleft
%  \begin{flushleft}
%\begin{center}
%\includegraphics[width=\linewidth]{e_dis_obsNew}
\includegraphics[width=1.15\linewidth]{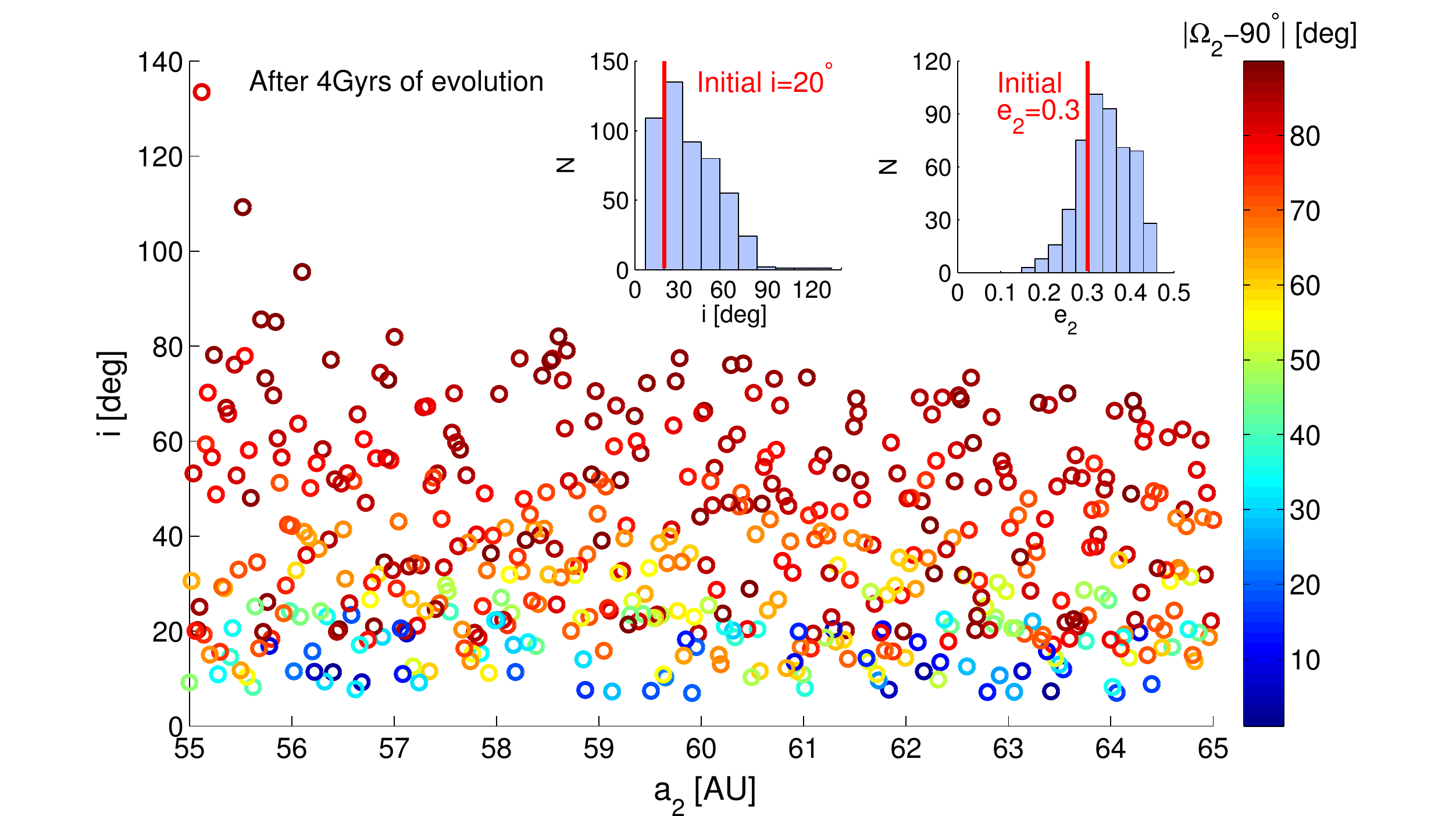}
\caption{\upshape {\bf Long timescale evolution of a debris disk.} The mutual inclination, $i$, as a function of the semi-major axis, $a_2$, after $4$~Gyrs of integration. The color code shows the longitude of ascending nodes $|\Omega_2-90^\circ|$ at that time (this presentation emphasizes the symmetry in the system). The right inset shows the histogram of the final eccentricity of the disk, while the left inset shows the histogram of the final mutual inclination of the disk.  As a proof of concept we consider a narrow debris disk located between $55-65$~AU, with an interior eccentric Jupiter ($e_1=0.85$) at $5$~AU around a solar mass star. The system was initially set with a mutual inclination of $20^\circ$, $e_2=0.3$ and $\Omega_2$ and $\omega_2$ are chosen from a random uniform distribution between $0-360^\circ$. {The results depicted here achieved by integrating over  the equations of motion, Eqs.~(\ref{eq:EOM1})-(\ref{eq:EOM4}). Note that GR effects are included here as well.} 
  } \label{fig:Puff}
   %\end{flushleft}
 % \end{center}
\end{figure}

 The system depicted in Figure \ref{fig:Kuip1} is, of course, just one example for a particular choice of the orbital parameters. To study the effects of planet's eccentricity and initial inclination we have systematically explored the $e_1$-$i$ parameter  space in Figure \ref{fig:TwoPanle_e5}. We choose a  Jupiter like system ($a_1=5$~AU) and for a range of eccentricities, with an outer orbit at  $55$~AU. The test particle orbit was initialized with $e_2=0.5$, $\Omega_2=90^\circ$, $\omega_2=0^\circ$ and a range of inclinations. 
% 
%  have systematically explored the $e_1$-$i$ parameter  space in Figure \ref{fig:TwoPanle_e5}, for a Jupiter like system ($a_1=5$~AU) and for a range of eccentricities. The outer orbit's set at $55$~AU was initially  set with $e_2=0.5$, $\Omega_2=90^\circ$ and $\omega_2=0^\circ$. 
  We show the maximum inclination reached during the evolution as a function of the initial inclination and eccentricity in Figure \ref{fig:TwoPanle_e5}. In the left panel we depict the initial inclination vs the initial Jupiter's eccentricity, where the color code marks the maximum inclination reached.  The solid line in the left panel follows  Equation  (\ref{eq:imaxmin}), which is consistent that the resonance associated with the quadrupole-level of approximation  is indeed the main driver for the dynamical evolution of the system. 
%  Of course the quadrupole-level is limited to below $90^\circ$ but the introduction of the octupole level (and GR by similar extent) allows the system to flip and becomes chaotic. 
  In the figure we depicted the  initial inclination regime to the prograde case ($i_{\rm initial}\leq 90^\circ$) to avoid clutter. However, we show in the right panel the maximum inclination reached during the evolution as a function of the initial inclination going all the way to $180^\circ$ this time.

As a proof of concept we depict in Figure \ref{fig:Puff} the behavior of a  narrow debris disk after $4$~Gyr of evolution. This inclination  represents the instantaneous inclination at $4$~Gyr of evolution. The system, of course, continue to oscillate, and the disk of particles will remain puffed. The inclination and eccentricity of the system at this snapshot are qualitatively different from the initial conditions assumed. 
 This  hypothetical system shows the orbital configuration of a disk located between $55-65$~AU, with an interior eccentric Jupiter ($e_1=0.85$) at $5$~AU around a solar mass star. The system was set initially with a mutual inclination of $20^\circ$, $e_2=0.3$ and $\Omega_2$ and $\omega_2$ are chosen from a random uniform distribution between $0-360^\circ$. While the particles in the disk have eccentric values, the disk does not appear as a coherent eccentric ring as the values of $\Omega_2$ and $\omega_2$ are random.   At the end of the integration, the particles in the disk became slightly more eccentric (with an average eccentricity of $\sim 0.34$) and there is a clear trend of $\Omega_2$ as a function of inclination. The particles with inclination above the initial $20^\circ$ have  a value of $\Omega_{2}$ close to zero (or with the symmetric value $180^\circ$), while the particles with inclination below $20^\circ$ have  $\Omega_2$ values closer to $90^\circ$. 
The behavior  is singular to  $\Omega_2$ and is not manifest itself in $\omega_2$ (does not depicted here to avoid clutter).  We note that some retrograde particles were formed as well, in line with \citet{Zanardi+17} numerical results of an orbital flip. These behaviors are easily understandable from the surface of section maps depicted above.

Its important to note that during the evolution of the system depicted in Figure \ref{fig:Puff}, the maximum $\epsilon$ achieved was $0.0625$. The  average value of $\epsilon$ which corresponds to the  maximum $e_2$ achieved during the evolution was $\sim 0.4$. Thus, the system is kept stable during this evolution, the secular approximation holds, and we do not expect any scattering event.  We also note that we have compared a debris disk particles secular and N-body evolution and found a qualitative agreement, which is similar in behavior to Figure \ref{fig:NB1} left panel, and this is not shown here to avoid clutter.

%  {\bf Gongjie, I didn't think we should show more ``proof of concept", such as circumbinary planet  or other things, as I think that this specific subject probably requires a study of its own and to add tides. What do you think?} \\
%  {\bf I agree that circumbinary planets and other things may require their own studies. Perhaps it is good to add a simple sentence? How about adding "This mechanism will also be important for circumbinary planetary systems, the stars near mergers of black hole binaries, but detailed studies is beyond the scope of this paper"?}

  \section{Discussion}\label{sec:dis}
  
  We have studied the secular evolution of  an outer test particle hierarchical system. We presented the three body, outer test body Hamiltonian up to the octupole level of approximation in the power series of the semi-major axis ratio.   We showed that in the quadrupole-level of approximation, $(a_1/a_2)^2$, the system has two distinct behaviors, librating and circulating (see Figure  \ref{fig:quad}), where the  librating mode gives the nominal precession of the nodes, results for which the inclination oscillates between the $i_{90}$ inclination (the inclination for which $\Omega_2=90^\circ$ and $180^\circ-i_{90}$). 
  %The two trajectory behaviors of the system is not limited to the quadrupole-level of approximation, as we find the 
  Furthermore, the bound values of the liberating mode have a simple analytical expression, Eq.~(\ref{eq:Omegamin}). We also found the minimum and maximum inclination that the system can reach in the circulating mode (see Equations  (\ref{eq:imaxmin}) and (\ref{eq:imim})). These conditions are sensitive to the initial inner orbit's eccentricity, and are nicely reproduced  in numerical testing here (see Figure \ref{fig:TwoPanle_e5}) and in \citet{Zanardi+17} numerical experiments (see their figure 14). We also estimated the timescale for oscillations for the two modes (see Section \ref{sec:Time}).
  
  We then showed that introducing the octupole level of approximation allows for transition between the two libration and circulation modes (see Figures \ref{fig:quadoct} and \ref{fig:Ex1}). This yields that the overall dynamics of the system is similar to the behavior of the general flip behavior in the eccentric Kozai-Lidov mechanism.  {In particular, the dynamics is quite chaotic for parameter regions with high $e_2$ and perpendicular mutual inclinations (when $J_2$ is low and when $J_{2,z}$ is near zero), as shown in the surface of sections Figures \ref{fig:sur1}-\ref{fig:sur3}.}

    General relativity can play an important role in suppressing or exciting the eccentricities in the hierarchical three body problem \citep[e.g.,][]{Naoz+12GR}. We find here similar behavior. Specifically, the {\it inclination} excitation will be suppressed for systems with GR precession faster than the quadrupole precession. However,  when GR  precession takes place on similar (or even somewhat larger) timescales to that of the quadrupole precession, the additional precession can produce  inclination excitations, in a non-regular manner (see Figure \ref{fig:GR}). 
  
  The dynamics of these type of systems can have a wide range of applications, from stars around supermassive black hole binaries to the evolution of individual debris disk particles. We have chosen the latter as an example and presented  a typical example of the evolution of a test particle due to the gravitational perturbations from an eccentric Jupiter (see Figure \ref{fig:Kuip1}). We systematically varied the Jupiter's eccentricity and the outer orbit's inclination, where we found an agreement between the analytical  relation for crossing the $90^\circ$ threshold,  and the numerical tests. This also suggests that  eccentric planet can pump up the inclination of icy body or comet reservoir analogs (Figure \ref{fig:TwoPanle_e5}). This was further supported by considering the evolution of an initially narrow, thin disk of test particles exterior to an eccentric planet, with initial mutual inclination of $20^\circ$. 
The disk became puffed with some particles on a retrograde orbits (see Figure \ref{fig:Puff}). A detailed study of the effects of eccentric planets on exterior test particle is presented in \citet{Zanardi+17}. 
  This mechanism will also be important for example to circumbinary planetary systems {\citep[considered first by][]{Ziglin75}} and the stars near mergers of black hole binaries, but detailed studies is beyond the scope of this paper.

%
%The architecture of a planetary system orbiting interior to a debris disk can leave a distinct imprint on the morphology of the disk. Recent observations found disks with exterior perturber or disks with large quantities of dust. The dust reported in many of these systems is expected to be short lived, and must either be transient or be continually replenished via collisions excited from interactions with larger bodies.
%
% represent are composed of rocky and icy material leftover from the formation of the star and any planets in the system.
%
% Eccentricity estimates  Taken from The Exoplanet Orbit Database and a sample with M sin i > 0.1MJ (Wright et al. 2011).
% 

\acknowledgements
We thank the referee for a quick and detailed report, and especially for his/her inquiry about the surface of sections. We also thank Vladislav Sidorenko for pointing out some missing references. 
SN acknowledges partial support from a Sloan Foundation Fellowship. GL is supported in part by the Harvard William F. Milton Award. MZ, GdE, and RPD acknowledge the financial support given by IALP, CONICET, and Agencia de Promoci\'{o}n Cient\'{\i}fica, through the PIP 0436/13 and PICT 2014-1292.

\appendix

\section{A. Orbital Parameters and the Scaled Time }\label{App:Ham}

One might have expected that $\omega_1 = {\rm const}$ when the outer particle is massless. But in truth $\Omega_1$ is undefined because the reference plane is aligned with the inner orbit. Therefore, the inner planet must only have $\omega_1+\Omega_1 = {\rm const}$, and we may choose without loss of generality the constant to equal zero. Hence, elimination of the nodes (i.e., $\Omega_1-\Omega_2=\pi$) implies $\omega_1=-\pi-\Omega_2$. 

%Note that Equation (\ref{eq:foct}) Can be written as:
%
%\begin{eqnarray}\label{eq:foct2}
%f_{\rm oct}  &= & \frac{ 15 e_1}{64 (1-e_2)^{3/2}} \bigg[     \frac{4+3e_1^2}{8}( 5\theta [3\theta -2]-1 )  \{ (1+\theta)\cos(\omega+\Omega) + (1-\theta)   \cos(\omega-\Omega) \}   \nonumber \\
%&+& \frac{35}{8} e_1^2 (1-\theta^2 ) \{ ( 1+\theta) \cos(\omega+3\Omega) +( 1-\theta) \cos(\omega-3\Omega) \} \bigg]
%\end{eqnarray}

%The relation between $f$ and the Hamiltonian is:
%\begin{equation}
%\Ham=\frac{k^2 m_1m_2}{a_1}\left( \frac{a_1}{a_2} \right)^3 f
%\end{equation}
%We note that $f_{\rm quad}$ has the same functional form as the inner test particle $F_{quad}$ presented in \citet{LN} up to the $(1-e_2)^{3/2}$ which is not constant in our case.
% and a factor of $1/3$.  

%We  rescale all momenta by an arbitrary constant without changing the equations of motion as long as we rescale the Hamiltonian by the same constant. Therefore we defined 
%\begin{equation}
%J=\frac{G_2}{m_2\sqrt{k^2 m_1 a_2}} = \sqrt{1-e_2^2}
%\end{equation}
%So the then the rescaled hamiltonian is:
%\begin{equation}
%F=\frac{\Ham}{m_2\sqrt{k^2 m_1 a_2}} =  \frac{P_2}{2\pi} \left( \frac{a_1}{a_2} \right)^2 f \ ,
%\end{equation}
%where $P_2$ is the period of the test particle around the star ($m_1$). 

Similarly to the treatment done in \citet{LN} we have rescaled the momenta by an arbitrary constant to achieve the specific angular momentum. The Hamiltonian is rescaled by the same constant, and we find that the rescaled time is:
\begin{equation}\label{eq:time}
\tau = \frac{t}{16} \frac{m_1 m_2}{(m_1+m_2)^2} \sqrt{\frac{G_N (m_1+m_2)}{a_2^3} }  \left( \frac{a_1}{a_2} \right)^2 =  \frac{ t}{16} \frac{m_1 m_2}{(m_1+m_2)^2} \frac{2\pi}{P_2}  \left( \frac{a_1}{a_2} \right)^2 \ ,
\end{equation}
where $t$ is the true time. The numerical factor $16$ comes by taking $m_3\to 0$ in the general Hamiltonian \citep[see][for the general form of the hierarchical three body double averaged hamiltonian]{Naoz16}. There is a choice to be made,  to either scale the Hamiltonian by this numerical factor or $\tau$. Here, we choose to absorb this number in $\tau$ to be consistent with the inner test particle Hamiltonian.

\section{B. Comparison with N-Body}\label{sec:Nbody}

\begin{figure*}
\hspace{-3cm}
\centering
\begin{minipage}[b]{.34\textwidth}
\centering
\includegraphics[scale=.45]{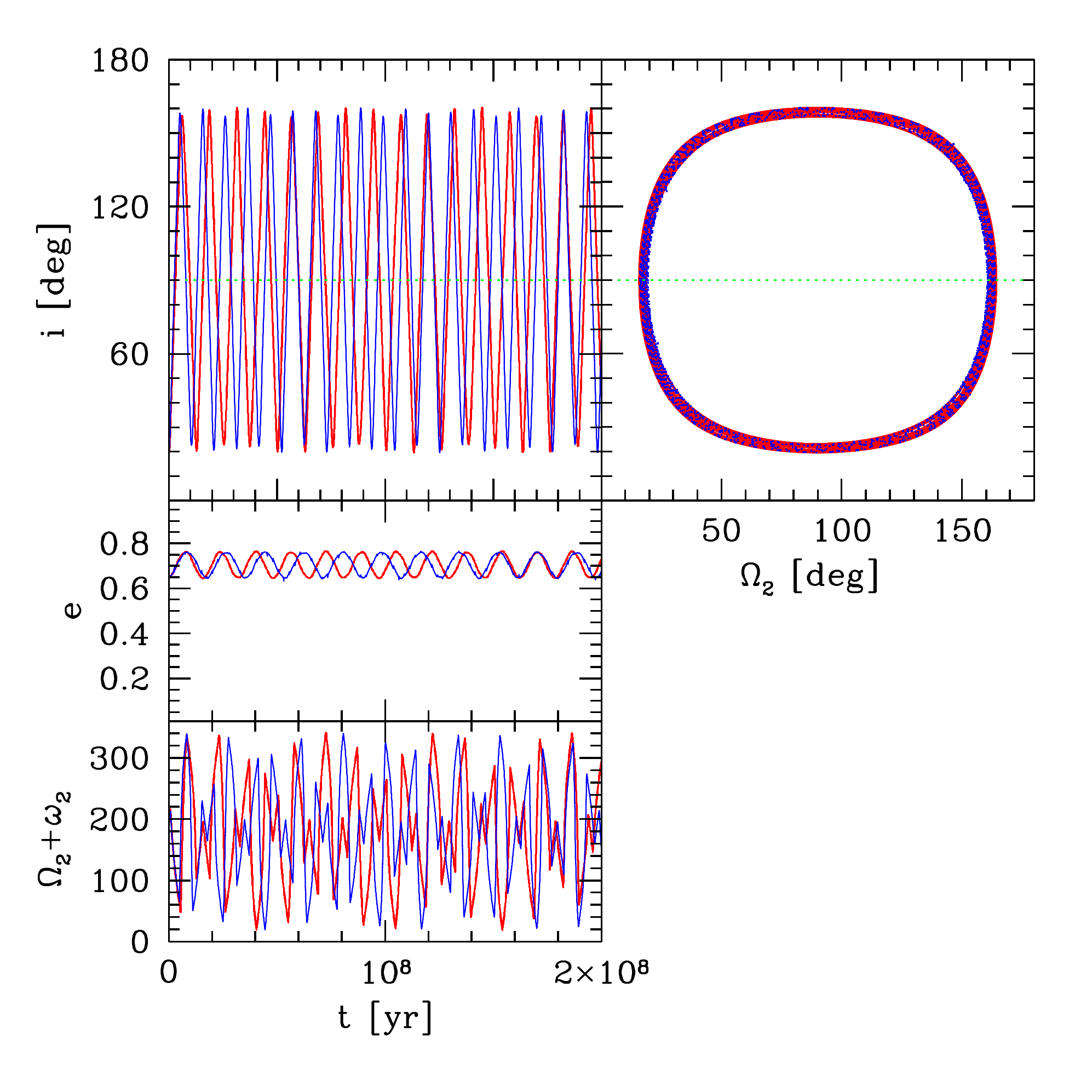}
\end{minipage}\hspace{3cm}%\qquad\qquad\qquad
\begin{minipage}[b]{.34\textwidth}
\centering
\includegraphics[scale=.45]{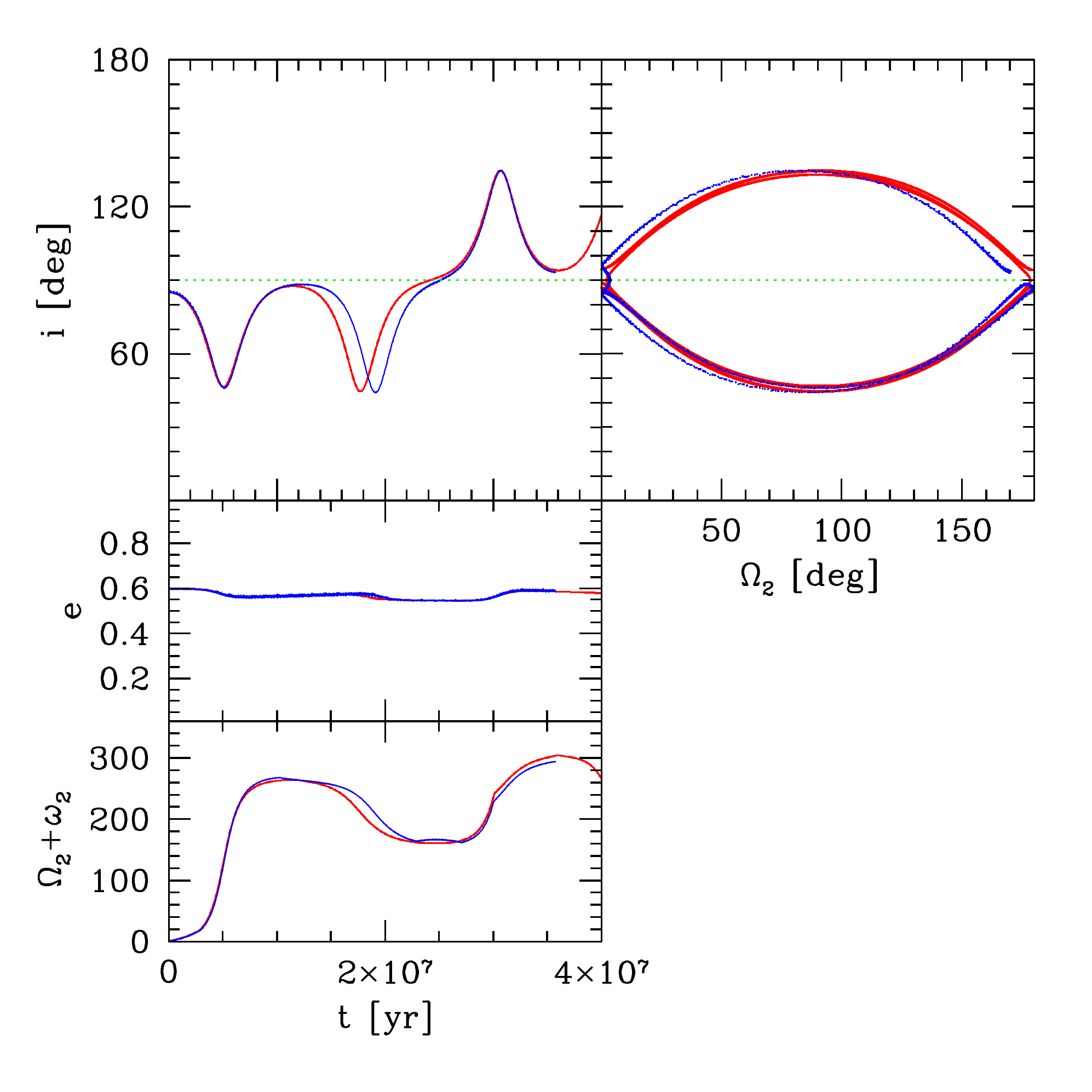}
\end{minipage}
\qquad\caption{ \upshape {\bf Comparison between the N-body results and the secular approximation at the octupole level.} We show the inclination, eccentricity and $\Omega_2+\omega_2$ as a function of time in the left hand side and $i-\Omega_2$ in the right hand panel. Red lines correspond to the secular calculation  (up to the octupole-level of of approximation) and blue lines corresponds to the N-body calculation. Note that in the $i-\Omega_2$ we depict the N-body results as points to allow for an easier comparison. {\bf On the left side} we consider the following system: $m_1=1$~M$_\odot$ $m_2=1$~M$_J$, $a_1=0.3$~AU, $a_2=40$~AU, $e_1=0.9$, $e_2=0.65$, $\omega_2=90^\circ$, $\Omega_2=100^\circ$, and $i=20^\circ$.  {\bf On the right side} we consider the following system: $m_1=1$~M$_\odot$ $m_2=1$~M$_J$, $a_1=0.5$~AU, $a_2=10$~AU, $e_1=0.4$, $e_2=0.6$, $\omega_2=\Omega_2=0^\circ$, and $i=85^\circ$. The evolution of this system was depicted in Figure \ref{fig:Ex1} and here we show this system for shorter evolution timescale to allow for a better comparison between the N-body and secular calculation.  }
%#1.     0.001    0.5    10.  0.4  0.6    0.      0.         85.          100.
\label{fig:NB1}
\hspace{1cm}
\end{figure*}
\begin{figure}
\begin{center}
\includegraphics[width=0.5\linewidth]{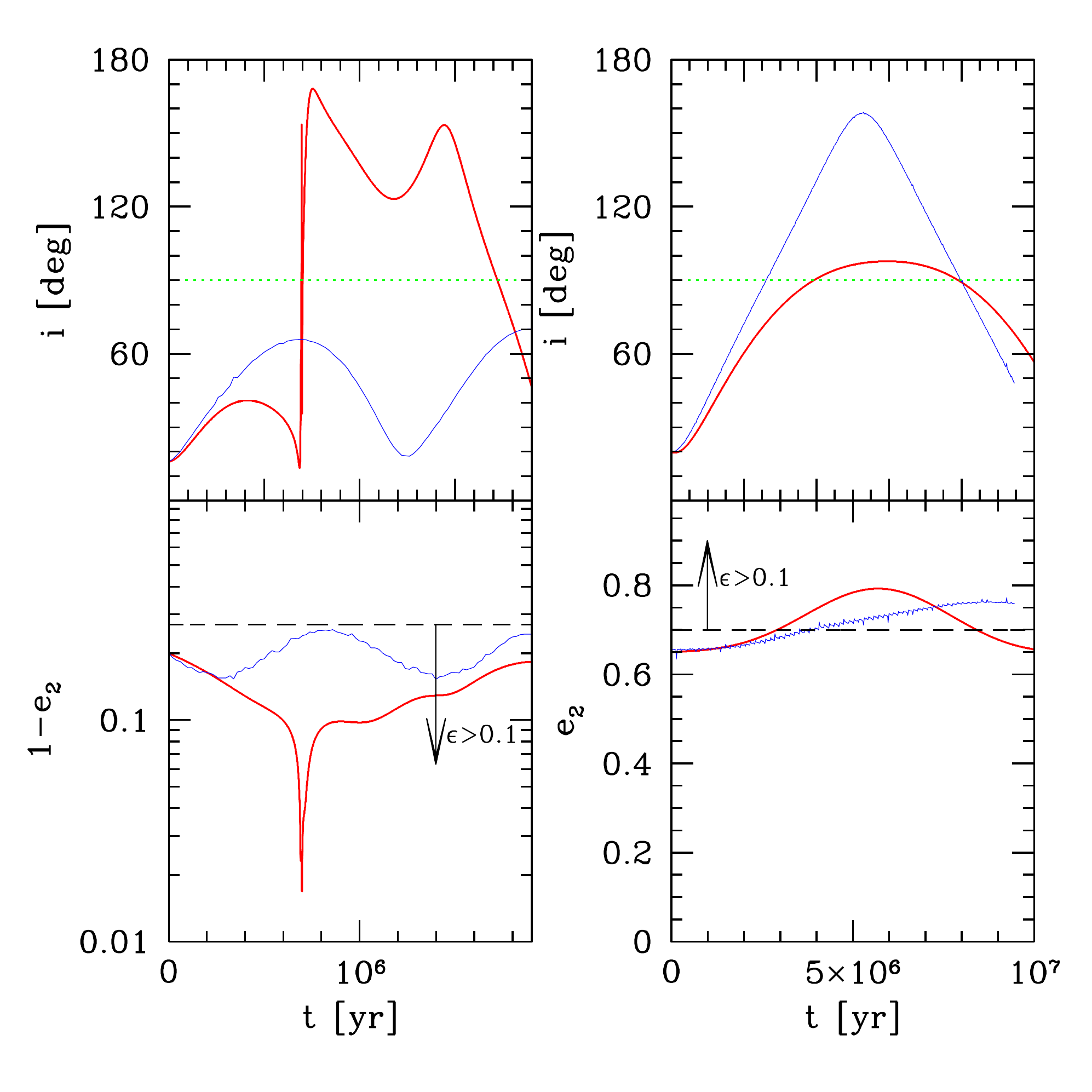}
\caption{\upshape {\bf Comparison between the N-body results and the secular approximation at the octupole level for systems around $\epsilon\sim 0.1$.} We show the inclination, eccentricity. Red lines correspond to the secular calculation  (up to the octupole-level of of approximation) and blue lines corresponds to the N-body calculation. {\bf Left side:} we consider the following system: $m_1=1$~M$_\odot$ $m_2=1$~M$_J$, $a_1=1$~AU, $a_2=15$~AU, $e_1=0.8$, $e_2=0.8$, $\omega_2=0^\circ$, $\Omega_2=90^\circ$, and $i=16^\circ$. These parameters imply an initial $\epsilon$ of $0.148$. {\bf Right side:} we consider the following system: $m_1=1$~M$_\odot$ $m_2=1$~M$_J$, $a_1=3$~AU, $a_2=40$~AU, $e_1=0.9$, $e_2=0.65$, $\omega_2=90^\circ$, $\Omega_2=100^\circ$, and $i=20^\circ$.  The latter systems initialize with $\epsilon=0.084$ however, as time goes by the eccentricity grows and the approximation breaks.
%1.     0.001    1.    15.  0.8  0.8    0.      -90.         16.          10.
%1.     0.001    3.    40.   0.9   0.65    90.      100.         20.          300.
} \label{fig:NB1F}
  \end{center}
\end{figure}

In this section, we compare the secular approximation at the octupole level in the test particle limit with the N-body simulation, using Mercury code \citep{Mercury}. In this comparison we did not include GR.   Good agreements can be reached when the apocenter distance of the inner binary is much smaller than the pericenter distance of the outer binary. In particular, we include an illustrative example here in Figure \ref{fig:NB1}, where we consider an eccentric ($e=0.9$ Jupiter at $3$~AU around a solar like object. The test particle is set at $40$~AU with $e_2=0.65$. The system is set initially with $i=20^\circ$,  $\omega_2=90^\circ$ and $\Omega_2=100^\circ$.

%%Specifically, we set the inner binary to be composed of a solar mass star and the Jupiter mass planet, with semi-major axis $a_1 = 0.3$AU and eccentricity $e_1 = 0.8$. For the outer companion, we set the mass to be a Pluto mass, and the semi-major axis to be $a2 = 10$AU. 
%
%\begin{figure*}
%\begin{center}
%\includegraphics[width=0.5\linewidth]{NBodyFail.pdf}
%\caption{\upshape {\bf Comparison between the N-body results and the secular approximation at the octupole level for systems around $\epsilon\sim 0.1$.} We show the inclination, eccentricity. Red lines correspond to the secular calculation  (up to the octupole-level of of approximation) and blue lines corresponds to the N-body calculation. {\bf Left side:} we consider the following system: $m_1=1$~M$_\odot$ $m_2=1$~M$_J$, $a_1=1$~AU, $a_2=15$~AU, $e_1=0.8$, $e_2=0.8$, $\omega_2=0^\circ$, $\Omega_2=90^\circ$, and $i=16^\circ$. These parameters imply an initial $\epsilon$ of $0.148$. {\bf Right side:} we consider the following system: $m_1=1$~M$_\odot$ $m_2=1$~M$_J$, $a_1=3$~AU, $a_2=40$~AU, $e_1=0.9$, $e_2=0.65$, $\omega_2=90^\circ$, $\Omega_2=100^\circ$, and $i=20^\circ$.  The latter systems initialize with $\epsilon=0.084$ however, as time goes by the eccentricity grows and the approximation breaks.
%%1.     0.001    1.    15.  0.8  0.8    0.      -90.         16.          10.
%%1.     0.001    3.    40.   0.9   0.65    90.      100.         20.          300.
%} \label{fig:NB1F}
%  \end{center}
%\end{figure*}

As shown in Figure \ref{fig:NB1}, the secular approximation (shown as red lines) agrees {\it qualitatively} well with the N-body results for the eccentricity and inclination oscillations. 
%However, the timescale seems to be shifted for the inclination variations. 
This is likely due to the double averaging process, but nonetheless, the maximum and minimum of the orbital parameters are conserved in both N-body and secular calculations. %truncation at the octupole level
% and the approximation of assuming the outer object to be a test particle.
It is interesting to note that similarly to \citep{LN} the approximation holds as long as $\epsilon<0.1$. However, unlike the inner-test particle case, $e_2$ can change and increase during the evolution which may break the validity of the approximation during the evolution, this is shown in Figures  \ref{fig:NB1F}.
%\bibliographystyle{apj}

%\newpage
\bibliographystyle{hapj}
%\bibliography{papers2}
\bibliography{Kozai}

\begin{thebibliography}{32}
\expandafter\ifx\csname natexlab\endcsname\relax\def\natexlab#1{#1}\fi

\bibitem[{{Blaes} {et~al.}(2002){Blaes}, {Lee}, \& {Socrates}}]{Bla+02}
{Blaes}, O., {Lee}, M.~H., \& {Socrates}, A. 2002, \apj, 578, 775,
  arXiv:astro-ph/0203370

\bibitem[{{Chambers} \& {Migliorini}(1997)}]{Mercury}
{Chambers}, J.~E., \& {Migliorini}, F. 1997, in Bulletin of the American
  Astronomical Society, Vol.~29, AAS/Division for Planetary Sciences Meeting
  Abstracts \#29, 1024--+

\bibitem[{{de la Fuente Marcos} {et~al.}(2015){de la Fuente Marcos}, {de la
  Fuente Marcos}, \& {Aarseth}}]{Marcos+15}
{de la Fuente Marcos}, C., {de la Fuente Marcos}, R., \& {Aarseth}, S.~J. 2015,
  \mnras, 446, 1867, 1410.6307

\bibitem[{{Farago} \& {Laskar}(2010)}]{Far+10}
{Farago}, F., \& {Laskar}, J. 2010, MNRAS, 401, 1189, 0909.2287

\bibitem[{{Ford} {et~al.}(2000){Ford}, {Kozinsky}, \& {Rasio}}]{Ford00}
{Ford}, E.~B., {Kozinsky}, B., \& {Rasio}, F.~A. 2000, \apj, 535, 385

\bibitem[{{Gallardo}(2006)}]{Gallardo06}
{Gallardo}, T. 2006, \icarus, 181, 205

\bibitem[{{Gallardo} {et~al.}(2012){Gallardo}, {Hugo}, \& {Pais}}]{Gallardo+12}
{Gallardo}, T., {Hugo}, G., \& {Pais}, P. 2012, \icarus, 220, 392, 1205.4935

\bibitem[{{Harrington}(1968)}]{Har68}
{Harrington}, R.~S. 1968, \aj, 73, 190

\bibitem[{{Harrington}(1969)}]{Har69}
------. 1969, Celestial Mechanics, 1, 200

\bibitem[{{Innanen} {et~al.}(1997){Innanen}, {Zheng}, {Mikkola}, \&
  {Valtonen}}]{Inn+97}
{Innanen}, K.~A., {Zheng}, J.~Q., {Mikkola}, S., \& {Valtonen}, M.~J. 1997,
  \aj, 113, 1915

\bibitem[{{Katz} {et~al.}(2011){Katz}, {Dong}, \& {Malhotra}}]{Katz+11}
{Katz}, B., {Dong}, S., \& {Malhotra}, R. 2011, ArXiv e-prints, 1106.3340

\bibitem[{{Kozai}(1962)}]{Kozai}
{Kozai}, Y. 1962, \aj, 67, 591

\bibitem[{Lee \& Chiang(2016)}]{Lee2016}
Lee, E.~J., \& Chiang, E.~I. 2016, The Astrophysical Journal, 1605.06118

\bibitem[{{Li} {et~al.}(2014{\natexlab{a}}){Li}, {Zhou}, \& {Zhang}}]{Li+14cir}
{Li}, D., {Zhou}, J.-L., \& {Zhang}, H. 2014{\natexlab{a}}, \mnras, 437, 3832

\bibitem[{{Li} {et~al.}(2014{\natexlab{b}}){Li}, {Naoz}, {Holman}, \&
  {Loeb}}]{Li+14Chaos}
{Li}, G., {Naoz}, S., {Holman}, M., \& {Loeb}, A. 2014{\natexlab{b}}, ArXiv
  e-prints, 1405.0494

\bibitem[{{Li} {et~al.}(2014{\natexlab{c}}){Li}, {Naoz}, {Kocsis}, \&
  {Loeb}}]{Li+13}
{Li}, G., {Naoz}, S., {Kocsis}, B., \& {Loeb}, A. 2014{\natexlab{c}}, \apj,
  785, 116, 1310.6044

\bibitem[{{Lidov}(1962)}]{Lidov}
{Lidov}, M.~L. 1962, planss, 9, 719

\bibitem[{{Lithwick} \& {Naoz}(2011)}]{LN}
{Lithwick}, Y., \& {Naoz}, S. 2011, \apj, 742, 94, 1106.3329

\bibitem[{Matthews {et~al.}(2014)Matthews, Krivov, Wyatt, Bryden, \&
  Eiroa}]{Matthews2014}
Matthews, B.~C., Krivov, A.~V., Wyatt, M.~C., Bryden, G., \& Eiroa, C. 2014, in
  Protostars and Planets VI, 521--544, arXiv:1401.0743v1

\bibitem[{{Murray} \& {Dermott}(2000)}]{MD00}
{Murray}, C.~D., \& {Dermott}, S.~F. 2000, {Solar System Dynamics}, ed.
  {Murray, C.~D.~\& Dermott, S.~F.}

\bibitem[{{Naoz}(2016)}]{Naoz16}
{Naoz}, S. 2016, \araa, 54, 441, 1601.07175

\bibitem[{{Naoz} {et~al.}(2011){Naoz}, {Farr}, {Lithwick}, {Rasio}, \&
  {Teyssandier}}]{Naoz11}
{Naoz}, S., {Farr}, W.~M., {Lithwick}, Y., {Rasio}, F.~A., \& {Teyssandier}, J.
  2011, \nat, 473, 187, 1011.2501

\bibitem[{{Naoz} {et~al.}(2013{\natexlab{a}}){Naoz}, {Farr}, {Lithwick},
  {Rasio}, \& {Teyssandier}}]{Naoz+11sec}
------. 2013{\natexlab{a}}, \mnras, 431, 2155, 1107.2414

\bibitem[{{Naoz} {et~al.}(2013{\natexlab{b}}){Naoz}, {Kocsis}, {Loeb}, \&
  {Yunes}}]{Naoz+12GR}
{Naoz}, S., {Kocsis}, B., {Loeb}, A., \& {Yunes}, N. 2013{\natexlab{b}}, \apj,
  773, 187, 1206.4316

\bibitem[{Nesvold \& Kuchner(2015)}]{Nesvold2015}
Nesvold, E.~R., \& Kuchner, M.~J. 2015, The Astrophysical Journal, 798, 83

\bibitem[{Nesvold {et~al.}(2016)Nesvold, Naoz, Vican, \& Farr}]{Nesvold2016}
Nesvold, E.~R., Naoz, S., Vican, L., \& Farr, W.~M. 2016, The Astrophysical
  Journal, 826, 1603.08005

\bibitem[{Rodigas {et~al.}(2014)Rodigas, Malhotra, \& Hinz}]{Rodigas2014}
Rodigas, T.~J., Malhotra, R., \& Hinz, P.~M. 2014, The Astrophysical Journal,
  780, 65

\bibitem[{{Teyssandier} {et~al.}(2013){Teyssandier}, {Naoz}, {Lizarraga}, \&
  {Rasio}}]{Tey+13}
{Teyssandier}, J., {Naoz}, S., {Lizarraga}, I., \& {Rasio}, F.~A. 2013, \apj,
  779, 166, 1310.5048

\bibitem[{{Verrier} \& {Evans}(2009)}]{Verrier+09}
{Verrier}, P.~E., \& {Evans}, N.~W. 2009, \mnras, 394, 1721, 0812.4528

\bibitem[{{Wright} {et~al.}(2011){Wright}, {Fakhouri}, {Marcy}, {Han}, {Feng},
  {Johnson}, {Howard}, {Fischer}, {Valenti}, {Anderson}, \&
  {Piskunov}}]{Wright+11}
{Wright}, J.~T. {et~al.} 2011, \pasp, 123, 412, 1012.5676

\bibitem[{{Zanardi} {et~al.}(2017){Zanardi}, {de El{\'{\i}}a}, {Di Sisto},
  {Naoz}, {Li}, {Guilera}, \& {Brunini}}]{Zanardi+17}
{Zanardi}, M., {de El{\'{\i}}a}, G.~C., {Di Sisto}, R.~P., {Naoz}, S., {Li},
  G., {Guilera}, O.~M., \& {Brunini}, A. 2017, ArXiv e-prints, 1701.03865

\bibitem[{{Ziglin}(1975)}]{Ziglin75}
{Ziglin}, S.~L. 1975, Soviet Astronomy Letters, 1, 194

\end{thebibliography}

\end{document}